\documentclass[10pt]{article}

\usepackage[utf8]{inputenc}  
\usepackage[T1]{fontenc}
\usepackage{lmodern}
\usepackage[english]{babel}  
\usepackage[a4paper,margin=1in]{geometry}
\usepackage{microtype}
\usepackage{graphicx}
\usepackage{booktabs}        

\usepackage{amsmath, amssymb}

\newcommand{\D}{\mathcal{D}}
\newcommand{\C}{\mathcal{C}}

\usepackage{siunitx}
\usepackage{enumitem}        
\usepackage{xcolor}
\usepackage{csquotes}
\usepackage{subcaption}
\usepackage{placeins}
\usepackage{pdflscape}
\usepackage{rotating}

\usepackage{mdframed}        

\newmdenv[
backgroundcolor=gray!10,
linecolor=gray!50,
skipabove=10pt,
skipbelow=10pt,
roundcorner=4pt,
leftmargin=0pt,
rightmargin=0pt
]{promptbox}

\usepackage{authblk}
\usepackage[hidelinks]{hyperref}
\hypersetup{
	colorlinks=true,
	linkcolor=blue!60!black,
	citecolor=blue!60!black,
	urlcolor=blue!60!black,
	pdfauthor={Isabella Mastroianni},
	pdftitle={Graph-Based Retrieval Strategy for RAG Systems in Hyperlinked Technical Documentation}
}
\usepackage{orcidlink} 

\title{\textbf{LARAG: Link-Aware Retrieval Strategy for RAG Systems in Hyperlinked Technical Documentation}}

\author[1]{Giorgia Bolognesi\orcidlink{0009-0003-8966-6211}}
\author[2]{Claudio Estatico\orcidlink{0000-0003-0948-6687}}
\author[3]{Ulderico Fugacci\orcidlink{0000-0003-3062-997X}}
\author[2,3]{Isabella Mastroianni\orcidlink{0009-0002-9866-3648}}
\author[1]{Claudio Muselli\orcidlink{0000-0002-1109-8125}}
\author[4]{Luca Oneto\orcidlink{0000-0002-8445-395X}}

\affil[1]{Rulex s.r.l., Genova, Italy}
\affil[2]{Department of Mathematics (DIMA), University of Genoa, Italy}
\affil[3]{Institute of Applied Mathematics and Information Technologies “Enrico Magenes” (IMATI), National Research Council, Italy}
\affil[4]{Department of Computer Science, Bioengineering, Robotics, and Systems Engineering (DIBRIS), University of Genoa, Italy}

\date{February 2026}

\begin{document}



\maketitle

\begin{abstract}
	\noindent
	
	Retrieval-Augmented Generation (RAG) enhances the factual grounding of Large Language Models by conditioning their outputs on external documents. However, standard embedding-based retrievers treat naturally structured corpora, such as technical manuals, as flat collections of passages, thereby overlooking the hyperlink topology that users rely on when navigating such content.
	
	We introduce LARAG (Link-Aware RAG): a lightweight, link-aware retrieval strategy that leverages the author-defined hyperlink structure already present in HTML documentation, encoding hyperlink relations as metadata in the chunk representations and exploiting them to perform a form of graph-like retrieval of locally relevant content.
	
	In a benchmark of twenty expert-designed queries over Rulex Platform technical documentation and four prompting strategies, LARAG consistently improves answer quality, achieving the highest BERTScore F1, while retrieving fewer chunks and generating fewer tokens than a baseline RAG architecture used for comparison. These results show that directly leveraging the existing hyperlink topology of technical documentation, even without explicit graph construction or inference, enables an implicit form of graph-like retrieval that yields a more faithful and efficient RAG pipeline, providing better grounding at lower cost.
	
\end{abstract}
\vspace{1.0em}

\vspace{2.0em}


\begin{center}
	\includegraphics[width=0.85\textwidth]{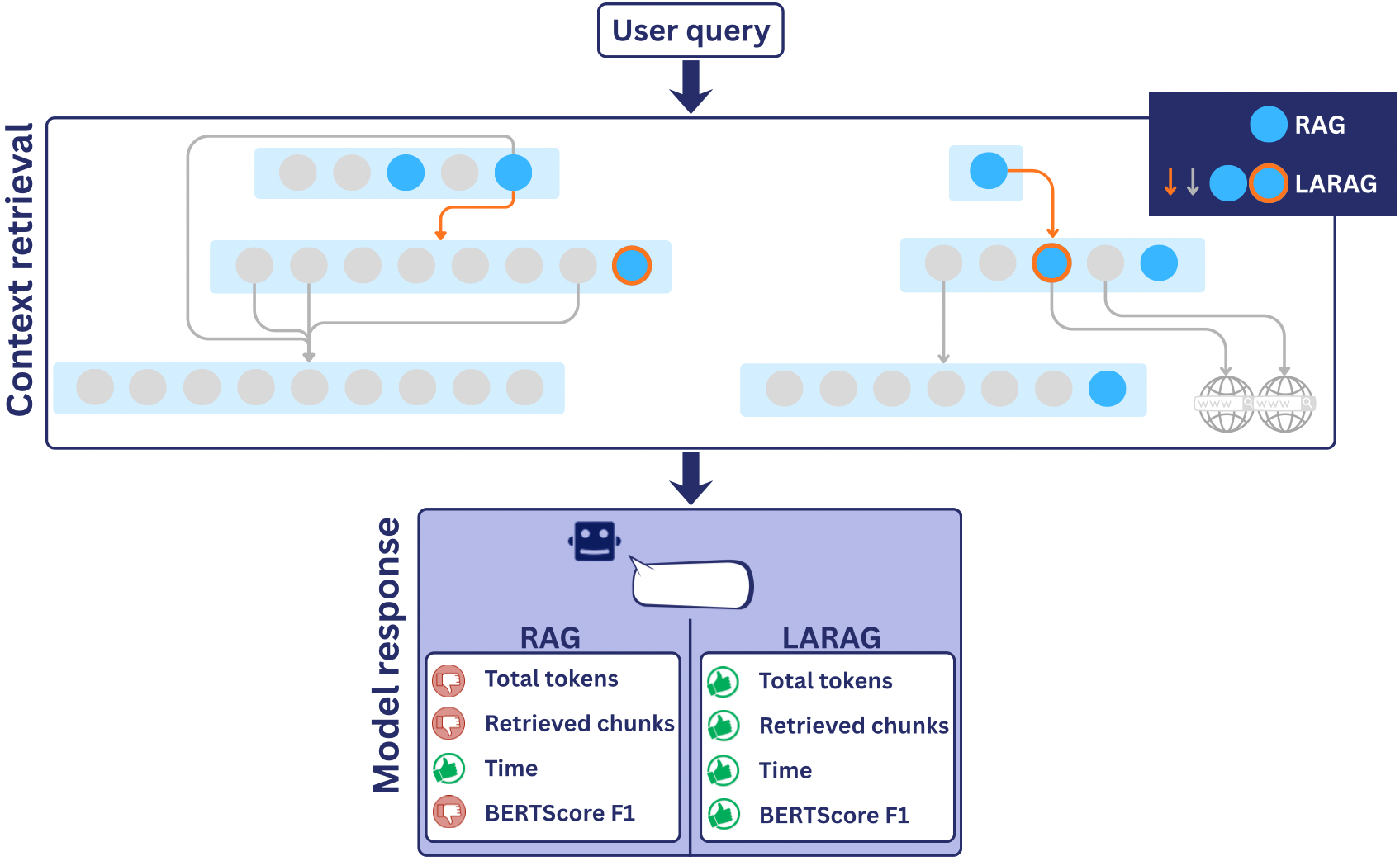}
\end{center}



\FloatBarrier
\section{Motivation and contributions}

\begin{sidewaysfigure}
	\includegraphics[width=\linewidth]{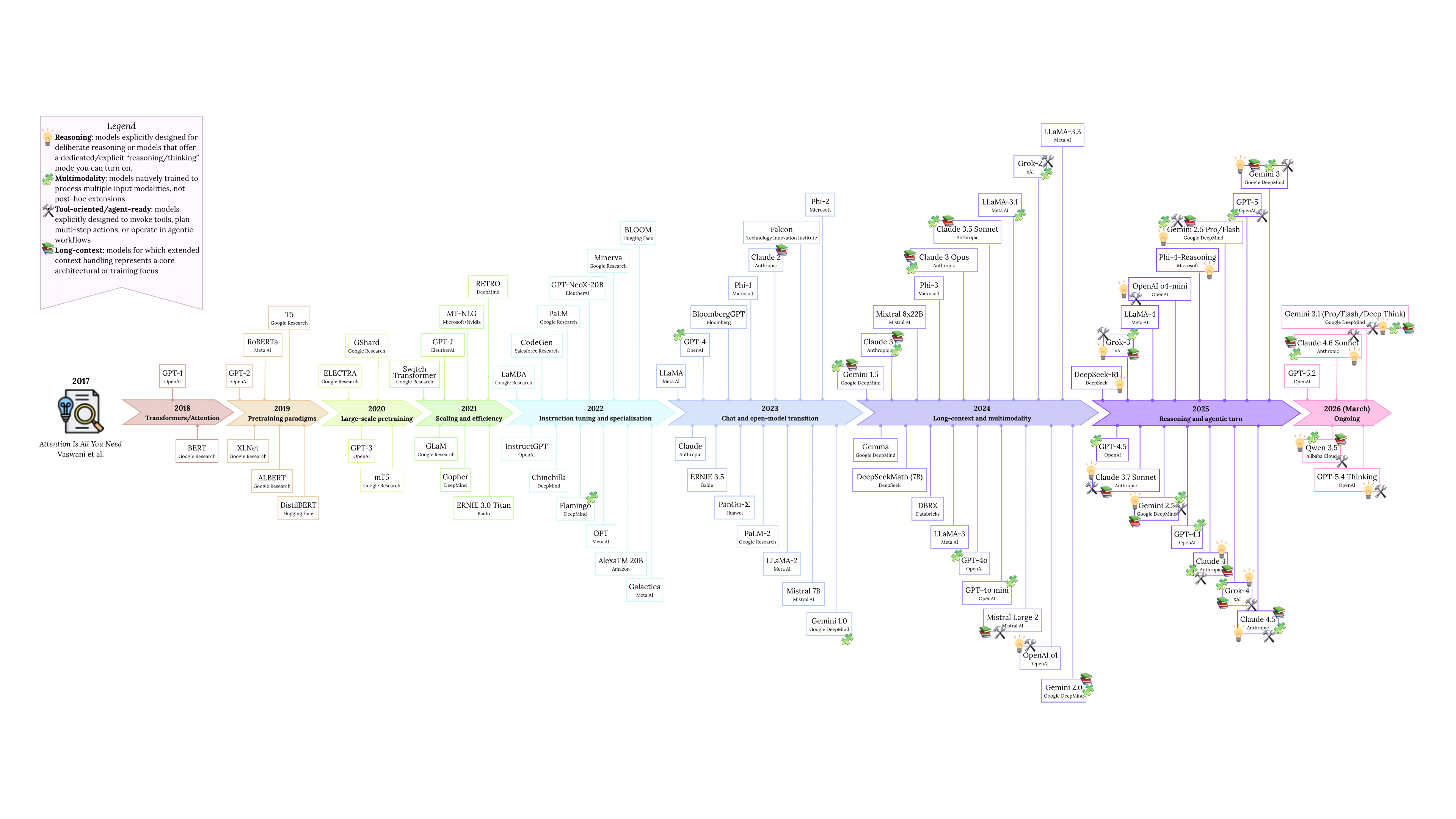}
	\caption{ 
		This diagram presents a simplified, panoramic view of key milestones in the evolution of LLMs, highlighting representative models and trends rather than providing a complete or strictly chronological account. evelopments shown for 2026 reflect an actively evolving phase and are based on direct usage and observed evolution, as technical documentation is not yet available. Icons are used only for unambiguous cases: since capabilities such as reasoning or tool use have blurred and evolving boundaries, icons are assigned only when a model can be unambiguously identified with a given capability. This diagram is inspired by a previously published timeline covering developments up to mid-2024 \cite{fi16100365}, and is here extended with more recent models as well as yearly paradigm labels and capability-based annotations.
	}
	\label{fig:evolution}
\end{sidewaysfigure}

Large Language Models (LLMs) have significantly improved the quality of natural language generation \cite{Xiao-and-Zhu:2025NLP,NEURIPS2020_1457c0d6, xiao2025foundations, fi16100365}, as evidenced by the steady increase in model scale, diversity, and adoption in recent years \cite{llama3, touvron2023llama, shao2024deepseekmath, guo2025deepseek, liu2024deepseek, agarwal2025gpt, floridi2020gpt3, openai2024gpt4technicalreport} (as well shown in Fig. \ref{fig:evolution}), yet their performance in knowledge-intensive tasks is still limited by the amount of information they can internally store and recall \cite{liu2024lost, hong2025context, hadi2023llm_survey, Huang_2025}. To address these limitations, several approaches have been explored, including full fine-tuning \cite{raffel2020exploring}, parameter-efficient methods such as LoRA \cite{hu2022lora}, and prompt-based techniques \cite{li2021prefix,lester2021power}. Retrieval-Augmented Generation (RAG) represents a complementary line of work \cite{lewis2020retrieval,asai2023self}, where external context is fetched at inference time.

By grounding generation in retrieved documents, RAG improves factual accuracy and reduces hallucinations. However, most RAG pipelines rely on embedding-based retrieval mechanisms that conceptualize the underlying corpus as a flat collection of independent passages. This assumption simplifies retrieval, but it discards structural and relational information that is often essential for navigating complex domains.

To overcome the limitations of flat retrieval, recent work has explored graph-aware or graph-based RAG extensions (GraphRAG), where documents, entities, or passages are connected through inferred or constructed relations \cite{knollmeyer2025document, dong2024don, hu-etal-2025-grag}. These approaches demonstrate that explicitly modeling relationships, often by constructing a graph and performing structured traversal or reasoning, can improve multi-hop retrieval and complex question answering. GraphRAG methods are therefore particularly well suited to scenarios where relational structure must be inferred from unstructured text or external knowledge bases.

However, technical documentation represents a distinct and underexplored case. Unlike general-purpose text, technical manuals are intentionally authored as hypertext, where sections, definitions, and procedures are explicitly interconnected via hyperlinks. These links encode the author’s intended conceptual dependencies and navigation paths, and are usually followed by human users to locate definitions, prerequisites, or related components. In this setting, much of the relevant structure is already present, explicit, and curated.
Despite this, existing RAG and GraphRAG approaches do not fully exploit this intentional hyperlink structure. Standard RAG pipelines ignore it altogether, while GraphRAG methods typically reconstruct relational structure through external graphs, entity linking, or learned representations, introducing additional complexity, overhead, and modeling assumptions that are often unnecessary in the presence of author-defined links. This mismatch motivates a different design choice: leveraging the existing hyperlink topology of technical documentation to guide retrieval.

The primary objective of our link-aware approach is therefore twofold. First, it aims to enable a more human-like navigation of technical documentation by explicitly exploiting hyperlink relations between sections. Second, it seeks to demonstrate that such structurally informed retrieval can generate accurate and faithful answers while requiring fewer tokens than a standard RAG pipeline. To evaluate these hypotheses, we directly compare our link-aware system LARAG with a baseline embedding-based RAG architecture, analysing whether the incorporation of hyperlink structure leads to measurable improvements in answer quality and retrieval efficiency.

\paragraph{Contributions.}
This paper makes the following contributions:
\begin{enumerate}
	\item We introduce \textbf{LARAG} (Link-Aware RAG), a \emph{link-aware retrieval strategy} that leverages the \emph{intentional hyperlink structure} of technical documentation.
	
	\item We show how hyperlink information can be used to support a form of graph-like retrieval and navigation without explicitly constructing or reasoning over a graph structure: \emph{hyperlink relations are exploited implicitly to guide localized expansion of relevant passages} within an otherwise standard and lightweight RAG pipeline, clearly distinguishing LARAG from existing GraphRAG approaches.
	
	\item We conduct an extensive evaluation on \emph{20 expert-level technical queries} across \emph{four different prompting strategies}, grounded in a real-world industrial documentation setting \cite{rulex_docs_14x}, directly comparing LARAG against a strong embedding-based RAG baseline \emph{under identical preprocessing conditions, isolating the effect of hyperlink-guided retrieval}.
	
	\item Beyond overall answer quality, we provide a \emph{fine-grained analysis} of precision, recall, token consumption, latency, and robustness with respect to document length, highlighting the efficiency-accuracy trade-offs introduced by link-aware retrieval.
\end{enumerate}

\FloatBarrier
\section{Related work}


\paragraph{Retrieval-Augmented Generation.}
Retrieval-Augmented Generation (RAG) enhances LLMs by conditioning generation on external context retrieved at inference time \cite{lewis2020retrieval}. Variants such as Self-RAG further couple retrieval and generation through iterative self-reflection \cite{asai2023self}. In all cases, the relevance and fidelity of generated answers depend critically on the retrieved context. Most RAG pipelines, however, assume that the underlying corpus can be treated as an unstructured set of passages, relying primarily on embedding-based similarity. While effective, this abstraction disregards document-level structure and inter-passage relationships that are often essential for navigating complex information spaces.

\paragraph{Structured and graph-based retrieval.}
Motivated by the limitations of flat retrieval, a growing body of work has explored structure-aware and graph-based extensions of RAG. Saad-Falcon et al.\ \cite{saad-falcon-etal-2024-pdftriage} show that document QA over structured PDFs benefits from retrieval methods that take layout and hierarchical organization into account. More generally, recent surveys describe the emergence of GraphRAG as a paradigm in which documents, entities, or passages are connected by inferred or constructed relations, enabling multi-hop retrieval and structured reasoning \cite{zhang2025survey,han2024retrieval,10.1145/3777378}.

Several approaches instantiate this idea by explicitly constructing graphs over text. UKRAG integrates structured and unstructured sources into a unified knowledge graph \cite{10.1007/978-3-031-82931-4_1}; From Local to Global introduces a GraphRAG architecture for query-focused summarization based on graph traversal \cite{edge2024local}; and Microsoft’s GraphRAG builds hierarchical indices to support multi-hop reasoning over private corpora \cite{microsoftgraphrag}. Other works derive graphs from textual structure or semantic representations: Knollmeyer et al.\ \cite{knollmeyer2025document} induce graphs from intrinsic document structure; Dong et al.\ \cite{dong2024don} construct AMR-based semantic graphs for reranking; and Hu et al.\ \cite{hu-etal-2025-grag} a pre-defined document graph by exploiting existing inter-document relations.


A related but more lightweight direction is the Graph RAG retriever available in LangChain \cite{langchaingraphrag}, which performs graph-guided retrieval by traversing relationships encoded in vector-store metadata. While this approach enriches similarity-based retrieval, the relations it operates on are externally provided metadata rather than links intrinsic to the source documents, and it does not model the author-defined structure of technical documentation.

Table~\ref{tab:contribution_vs_prior} summarizes these differences by positioning LARAG with respect to representative RAG and GraphRAG approaches along key design dimensions.

%

\paragraph{Prior work.}
This work builds on the hyperlink-aware retrieval strategy first introduced by Bolognesi~\cite{bolognesi2025thesis}, which explored enriching RAG with internal documentation hyperlinks and reported strong empirical results on a seven-query benchmark using four prompt strategies. In that study, the baseline RAG and the link-aware system were implemented with different preprocessing pipelines, including distinct chunking strategies and metadata representations, reflecting a broader architectural comparison.

In contrast, our work adopts a controlled experimental setup in which RAG and LARAG share the same preprocessing (i.e., identical parsing, chunking strategy, chunk size, and overlap), differing only in the activation of hyperlink-guided traversal. This setting allows us to attribute the observed performance gains unambiguously to link-aware retrieval itself. Moreover, we substantially broaden and refine the analysis by adopting a larger benchmark of twenty expert-level queries (while maintaining the same four prompt strategies), evaluating all components of BERTScore (F1, Precision, and Recall), analyzing the impact of reference and prediction length, and investigating the relationship between semantic quality and computational cost in terms of token usage and latency. Together, these extensions provide a more comprehensive and systematic assessment of link-aware retrieval for technical documentation.

\begin{table*}[!h]
	\centering
	\footnotesize
	\setlength{\tabcolsep}{6pt}
	\begin{tabular}{l p{3.2cm} p{1.3cm} p{2cm} p{1.8cm} p{1.8cm}}
		\toprule
		\textbf{Method} &
		\textbf{Edges origin} &
		\textbf{Explicit graph} &
		\textbf{Traversal cost} &
		\textbf{Dependence on learned edges} &
		\textbf{Support for authored links} \\
		\midrule
		
		
		Knollmeyer et al.~\cite{knollmeyer2025document} &
		Intrinsic structure &
		Yes & Low & No & No \\
		
		Dong et al.~\cite{dong2024don} &
		AMR parsing &
		Yes & Low--Med & Yes & No \\
		
		UKRAG~\cite{10.1007/978-3-031-82931-4_1} &
		External + corpus &
		Yes & High & Yes & No \\
		
		Edge et al.~\cite{edge2024local} &
		Induced relations &
		Yes & High & Yes & No \\
		
		Microsoft GraphRAG~\cite{microsoftgraphrag} &
		Clustering / summaries &
		Yes & High & Yes & No \\
		
		GRAG~\cite{hu-etal-2025-grag} &
		Induced semantics &
		Yes & Med--High & Yes & No \\
		
		LangChain Graph RAG \cite{langchaingraphrag} &
		User-provided &
		No & Low & No & No \\
		
		\textbf{LARAG (ours)} &
		\textbf{HTML internal links} &
		\textbf{No} & \textbf{Low} & \textbf{No} & \textbf{Yes} \\
		
		\bottomrule
	\end{tabular}
	\caption{
		Comparison between LARAG and representative GraphRAG approaches.
	}
	\label{tab:contribution_vs_prior}
\end{table*}

\FloatBarrier
\section{Problem formalization and proposed approach}




Let $M$ be a Large Language Model and let $\D$ be a hyperlinked corpus structured into sections connected through explicit anchors and links.

\vspace{5pt}
\noindent\textbf{Problem.} Given a query $q$ in natural language, retrieve a context $\C \subseteq \D$ that enables $M$ to maximize both the faithfulness and the completeness of the answer produced in response to $q$.
\vspace{5pt}

This formulation does not prescribe how $\C$ should be selected, leaving open whether $\D$ ought to be treated as an unstructured collection of passages or as a hypertext with an explicit topology. 
These alternative interpretations correspond to different retrieval paradigms: standard RAG, which ignores document-level structure; graph-based approaches such as GraphRAG, which explicitly construct or induce a graph over the corpus; and our proposed method, LARAG, which leverages the hyperlink topology already provided by the documentation without constructing a separate graph representation.

\medskip
To assess the impact of leveraging this topology, we compare these two perspectives empirically through a baseline RAG system and our link-aware LARAG architecture.


\paragraph{RAG.}
A classical RAG system can be described as a triple $(\D, R, M)$, where $\D$ is treated as a flat collection of independent passages, $R$ is a retriever, and $M$ is a generative model. Given a query $q$, the system produces an answer through two main phases
\begin{enumerate}
    \item \textit{Retrieval phase}. The retriever $R$:
        \begin{itemize}
            \item assigns each $d \in \D$ a relevance score $R(q,d)$, based on embedding similarity with $q$;
            \item selects the top-$k$ documents, forming a context $\C = \{d_1,\dots,d_k\}$.
        \end{itemize}
    \item \textit{Generation phase}. The model $M$ then generates an answer $M(q,\C)$, whose faithfulness and completeness depend solely on the semantic match between $q$ and the selected passages.
\end{enumerate}

This baseline formulation treats $\D$ as an unstructured set. We therefore introduce a link‑aware variant that explicitly leverages the document-level topology encoded in hyperlinks.

\paragraph{LARAG.} 
As in the baseline case, the system is described by the triple $(\D, R, M)$. However, in LARAG the corpus $\D$ is viewed through the navigational structure induced by its hyperlinks. Hyperlink relations are encoded as metadata associated with each chunk, enabling the retriever to access the author-defined structure of the documentation without maintaining a separate graph representation, providing a lightweight form of graph traversal
.

The overall pipeline still operates through the same two phases as before, but the retriever $R$ adopts a link-aware retrieval strategy consisting of two steps:

\begin{enumerate}[label=1.\alph*.]
	\item \textit{Initial retrieval}. As in the baseline system, the retriever $R$ computes an embedding-based relevance score $R(q,d)$ for each $d \in \D$, and selects the top-$k$ documents to form an initial context $\C_0 = \{d_1,\dots,d_k\}$.
	\item \textit{Link-based expansion and reranking}. Starting from each $d_i \in \C_0$, the system retrieves any documents whose hyperlinks are listed in the metadata of $d_i$, thereby collecting additional candidates that are topologically related to the initially retrieved passages. The union of these candidates and the initial retrieval results is then reranked according to semantic similarity with the query, after which the top-$k$ elements are selected to form the final context $\C$ supplied to the generator.
\end{enumerate}

The resulting context $\C$ integrates both semantic relevance and structural coherence, enabling $M$ to access passages that are not only lexically aligned with the query but also connected through the hyperlink structure intentionally defined by the authors of the documentation, as one can appreciate in the example shown in Figure \ref{fig:k5vsk10vsLA}.

By summarizing the rationales, LARAG and GraphRAG approaches share the high-level goal of exploiting relational structure during retrieval, but they are grounded in fundamentally different assumptions. GraphRAG methods are designed for scenarios in which relational structure must be explicitly constructed or induced from the corpus, e.g., via knowledge graphs, semantic parsing, clustering, or learned inter-document relations, and therefore rely on the materialization and traversal of an explicit graph representation. LARAG instead targets a complementary setting, typical of technical documentation, where much of the relevant structure is already explicit and intentionally authored. Rather than inducing a graph, LARAG directly leverages existing HTML hyperlinks, encoding them as metadata and exploiting them implicitly at retrieval time. This enables hyperlinks navigation without constructing, storing, or reasoning over a standalone graph object.

Table~\ref{tab:contribution_vs_prior} positions LARAG with respect to representative GraphRAG variants along key design dimensions, including the origin of relational edges, the presence of explicit graph construction, traversal cost, and support for authored links. While the focus of this work is on isolating the impact of author-defined hyperlinks, combining such signals with graph-based retrieval methods represents an interesting direction for future investigation, as discussed in Section \ref{sec:conclusions}.




\section{Methodology}
This section describes the methodological framework adopted in this work, outlining the tools and infrastructure used, the retrieval architectures evaluated, and the prompt design strategies explored in our experiments.

\FloatBarrier
\subsection{Infrastructure and tools}
This work relies on a variety of technologies, software tools, and libraries. The selection of each component was driven by the need for computational efficiency, compatibility with LLM-based architectures, and alignment with current research practices. In particular, the implementation required tools for managing and partitioning textual data, performing semantic search through vector embeddings, and interacting with generative models via Application Programming Interfaces (APIs). These technologies form the common foundation shared by both retrieval approaches developed in this project.

All components were implemented in Python using standard tools for LLM-based retrieval and generation. LangChain was used to manage document processing, vector-store interaction, and prompt orchestration. Chroma served as the vector database for storing and retrieving embeddings. We compared several OpenAI embedding models on MIRACL and MTEB benchmarks and selected \texttt{text-embedding-3-large} for its superior average performance. For answer generation, we employed the \texttt{gpt-4o-mini} model, accessed through Azure OpenAI endpoints. Cosine similarity was adopted as the scoring metric for retrieval due to its efficiency and scale-invariance under unit-normalized embeddings.

Table \ref{tab:tech-overview} summarizes the technologies and their respective functions within the systems.

\begin{table*}[t]
\centering
\small
\renewcommand{\arraystretch}{1.2}
\begin{tabular}{p{3.4cm} p{3.4cm} p{7.8cm}}
\toprule
\textbf{Tool / Library} & \textbf{Category} & \textbf{Role in the Project} \\
\midrule

Python & Programming Language &
Implemented the entire retrieval and evaluation pipeline. \\

LangChain & Framework &
Managed document processing, prompt handling, and integration with external components. \\

Chroma DB & Vector Database &
Stored and retrieved dense document embeddings for semantic search. \\

\texttt{text-embedding-3-large} & Embedding Model &
Converted textual passages into semantic vector representations. \\

\texttt{gpt-4o-mini} & Language Model &
Generated answers conditioned on the retrieved context. \\

Azure OpenAI & Cloud Service &
Provided scalable and secure access to GPT and embedding models. \\

Cosine Similarity & Similarity Metric &
Computed similarity scores between embedding vectors during retrieval. \\

\bottomrule
\end{tabular}
\caption{Overview of the main technologies and their roles within the project.}
\label{tab:tech-overview}
\end{table*}


\FloatBarrier
\subsection{RAG architectures}

Both systems evaluated in this work follow the standard Retrieval-Augmented Generation (RAG) pipeline, consisting of six main stages:
\begin{enumerate}[label=\bfseries(\roman*)]
	\item document parsing,
	\item chunking,
	\item embedding and indexing,
	\item retrieval,
	\item prompt augmentation,
	\item answer generation.
\end{enumerate}
The two architectures share the same overall workflow and models; they differ only in how document structure is interpreted and exploited, particularly during parsing, chunking, and retrieval. Below, we describe each pipeline step and highlight the differences between RAG and LARAG.

\FloatBarrier
\subsubsection{Document preprocessing and database creation}

Here we describe the first part of the two pipelines, i.e. passages \textbf{(i)}--\textbf{(iii)}. In both the pipelines, after chunking, each text segment is converted into a \texttt{Document} object containing the chunk content and its associated metadata. While both architectures rely on the same embedding model and vector database, they differ substantially in the structure and richness of their metadata.

\begin{figure}[!h]
	\centering
	\includegraphics[width=0.75\textwidth]{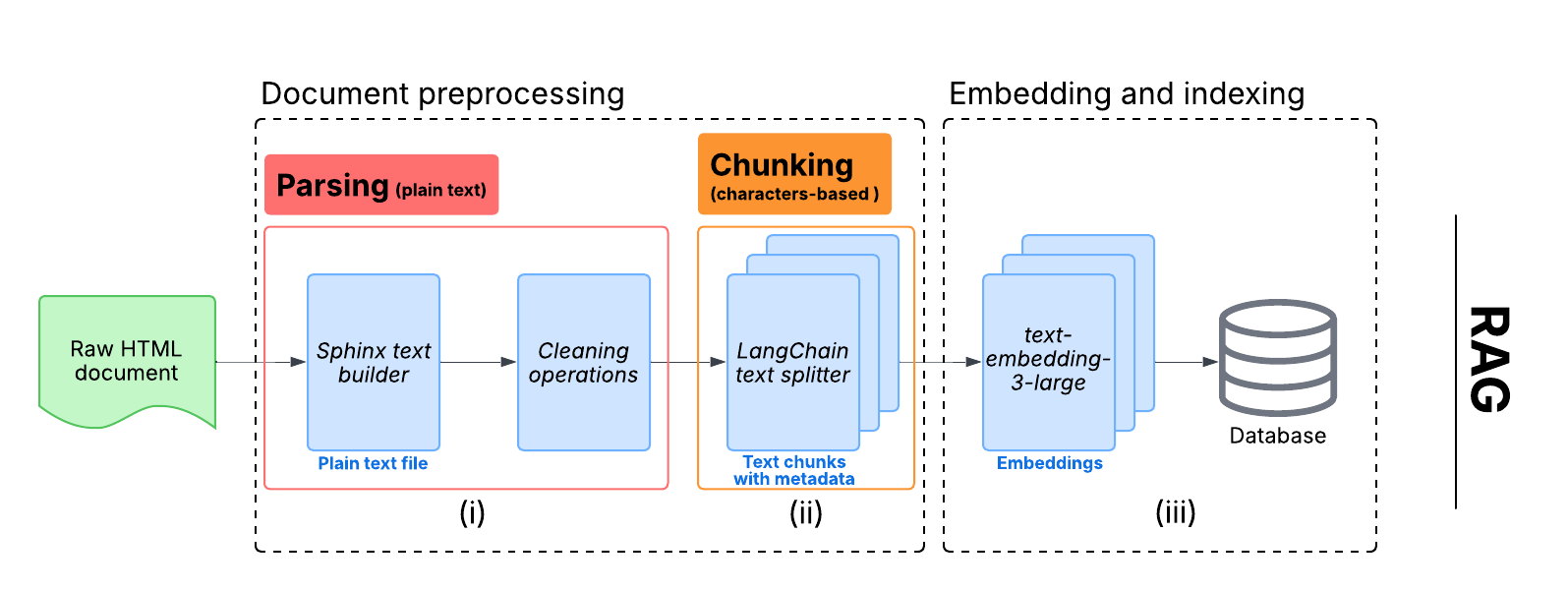}
	\includegraphics[width=\textwidth]{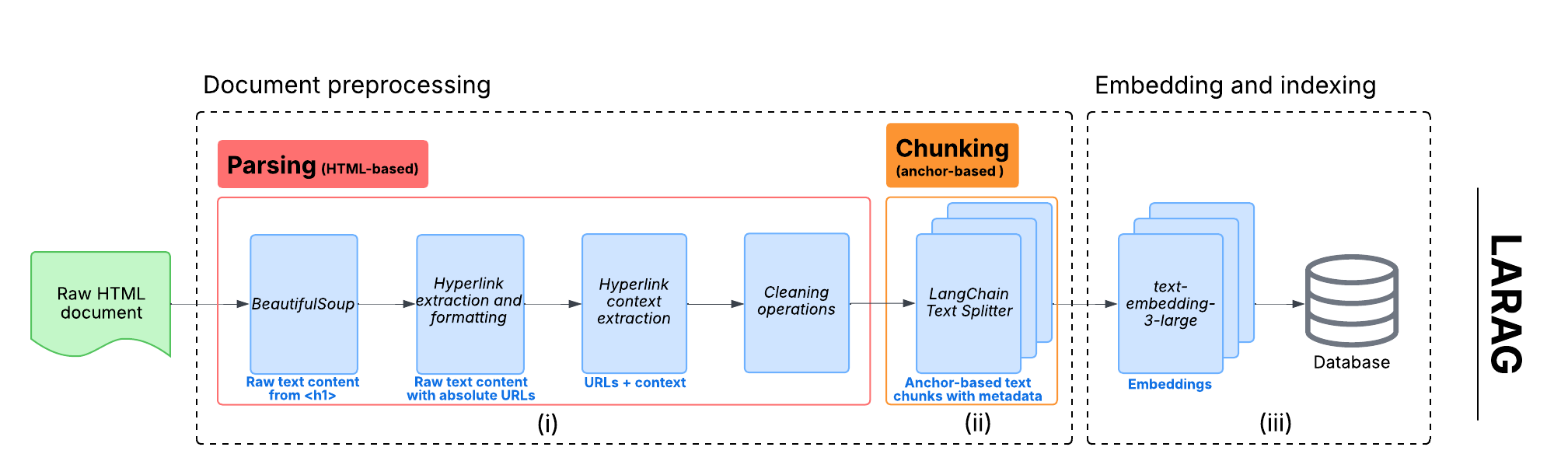}
	\caption{Preprocessing pipelines for RAG (top) and LARAG (bottom).}
	\label{fig:preproc}
\end{figure}

\paragraph{RAG.} \textbf{(i)} Documentation files are exported in plain-text format, which removes HTML structure and discards all hyperlinks and anchors. Minimal metadata (source file and identifier) are preserved (see Fig. \ref{fig:preproc}, top). \textbf{(ii)} A recursive character-based splitter produces chunks of 800 characters with an overlap of 100, yielding flat segments that do not reflect section structure (see Fig. \ref{fig:preproc}, top). This configuration was selected following empirical testing in order to reduce information loss at chunk boundaries. In particular, concepts occurring at the end of a chunk, such as references to subsequent sections or introductory sentences, are duplicated in the following chunk, ensuring greater semantic cohesion and improving the model’s understanding during the retrieval phase. 
This design choice is especially relevant for technical documentation, where key concepts frequently span multiple sentences.
Each chunk is stored with minimal metadata, including only:
\begin{itemize}[leftmargin=*]
	\item \texttt{source}: the absolute path of the original document,
	\item \texttt{id}: a unique identifier in the form \texttt{<source>:<chunk\_index>}.
\end{itemize}

Example:
\begin{promptbox}
	\begin{verbatim}
		Document(
		page_content = "... Rulex Studio Homepage  Before ...",
		metadata = {
			"source": "/v14/studio/index.txt",
			"id": "/v14/studio/index.txt:2"
		}
		)
	\end{verbatim}
\end{promptbox}

\paragraph{LARAG.} \textbf{(i)} Files are parsed directly in HTML format using \texttt{BeautifulSoup},\footnote{\href{https://beautiful-soup-4.readthedocs.io/en/latest/}{https://beautiful-soup-4.readthedocs.io/en/latest/}} preserving anchors, section boundaries, and all internal hyperlinks. Each link is normalized to an absolute URL and stored together with a short surrounding linguistic context (twelve words), providing semantic cues on the relationship between pages (see Fig. \ref{fig:preproc}, bottom). \textbf{(ii)} Chunking follows Sphinx-generated anchors,\footnote{Sphinx is a documentation generator that transforms structured text sources (e.g.\ reStructuredText or Markdown) into multiple output formats such as HTML and PDF. During this process, it automatically creates a hierarchical structure with anchors, cross-references, and indexes. The Rulex Platform documentation is authored and built using Sphinx.} producing semantically coherent sections aligned with HTML hierarchy. These sections are then subdivided using a recursive splitter with a chunk size of 1{,}000 characters and a 150-character overlap. While overlap is also employed in the baseline RAG pipeline, larger chunk sizes and overlaps are adopted in LARAG to better match the structure of the HTML documentation, which is organized into more coherent and information-dense sections. Increasing the chunk size helps avoid excessive fragmentation of content belonging to the same thematic unit, while a larger overlap preserves contextual continuity across chunk boundaries. This is particularly important in the presence of internal references and logical transitions between sections, which are common in technical documentation and critical for hyperlink-guided retrieval. Metadata include: source URL, anchor name, unique identifier, list of outgoing links, and link contexts (see Fig. \ref{fig:preproc}, bottom). In the link-aware architecture, metadata encode the hypertextual structure extracted from the HTML. Each \texttt{Document} includes:
\begin{itemize}[leftmargin=*]
	\item \texttt{source}: absolute URL of the page containing the chunk,
	\item \texttt{anchor\_name}: the Sphinx-generated anchor associated with the section,
	\item \texttt{id}: a unique identifier combining URL and anchor: \texttt{<source>:<anchor\_name>-<chunk\_number>},
	\item \texttt{links}: the list of outgoing hyperlinks,
	\item \texttt{links\_context}: short linguistic contexts (twelve words) surrounding each hyperlink in the source HTML.
\end{itemize}

Example:
\begin{promptbox}
	\begin{verbatim}
		Document(
		page_content = "... Rulex Studio Homepage  Before ...",
		metadata = {
			"source": "https://doc.rulex.ai/docs/v14/studio/index.html",
			"anchor_name": "rulex-studio-homepage",
			"id": "https://doc.rulex.ai/docs/v14/studio/index.html:rulex-studio-homepage-1",
			"links": [
			"https://doc.rulex.ai/docs/v14/platform/index.html#platform-overview",
			...
			],
			"links_context": [
			"If some concepts are not clear for you please refer to the platform section ..."
			]
		}
		)
	\end{verbatim}
\end{promptbox}

These richer metadata support hyperlink-aware retrieval by exposing the document topology that is lost in the plain-text baseline.

\medskip

In both pipelines, step \textbf{(iii)} embeds all \texttt{Document} objects using \texttt{text-embedding-3-large} and stores them in a ChromaDB index, relying on the same embedding model and storage backend for both architectures. The resulting metadata, however, differ depending on whether the RAG or the LARAG pipeline is applied. Duplicate entries are prevented by checking each document's unique identifier (see Fig. \ref{fig:preproc}).

\FloatBarrier
\subsubsection{Context retrieval and response generation}

Here we describe the second part of the two pipelines, corresponding to steps \textbf{(iv)}–\textbf{(vi)}, which cover context retrieval and final answer generation.

\begin{figure}[!ht]
	\centering
	\includegraphics[width=0.55\textwidth]{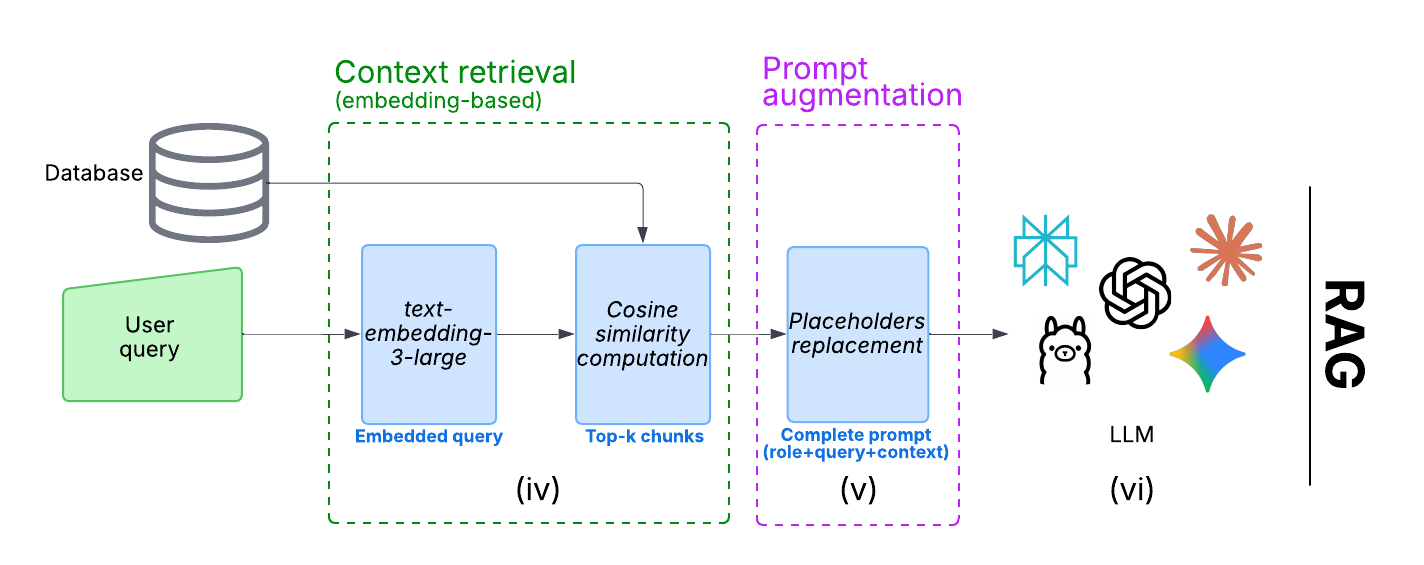}
	\includegraphics[width=\textwidth]{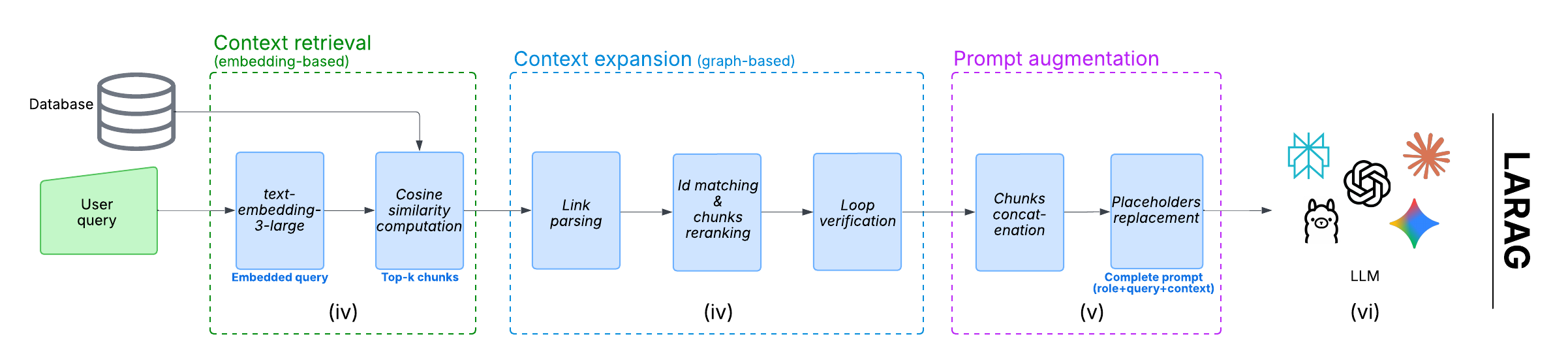}
	\caption{Context retrieval pipeline for RAG (top) and LARAG (bottom).}
	\label{fig:retrieval}
\end{figure}

\paragraph{RAG.} \textbf{(iv)} The query $q$ is embedded and cosine similarity is computed against all stored vectors. The top-$k$ most similar chunks (with $k=5$ or $k=10$, see Section \ref{subsec:optimal configuration}) form the context $\C$ (see Fig. \ref{fig:retrieval}, top).

\paragraph{LARAG.} 
\textbf{(iv)} Retrieval begins with the same initial semantic step, yielding a seed set $\mathcal{C}_0$. From this set, the retriever performs a hyperlink-aware expansion process. Importantly, this does not operate over an explicit graph structure: hyperlinks are not materialized through a stored graph. Instead, outgoing links are accessed directly from the metadata of each retrieved chunk and traversed on-the-fly using a controlled depth-first search (DFS), parameterised by \texttt{n\_links}, \texttt{depth}, and \texttt{top\_m} (see Section~\ref{subsec:optimal configuration}). 
The DFS hyperlink-based expansion operates according to the following principles:

\begin{itemize}
	
	\item \textbf{Unit of expansion.}  
	Hyperlinks target documentation sections, that have been further divided into chunks during preprocessing. Expansion is thus performed at the chunk level: for each link, candidate chunks from the target section are evaluated and selected via the context-based reranking mechanism explained below. 
	
	\item \textbf{Cycle avoidance and duplicate handling.}  
	The expansion maintains a global set of visited chunks identifiers shared across the entire DFS traversal, including expansions originating from different seed chunks. Each chunk is expanded at most once: if a candidate target chunk has already been visited, it is skipped. This mechanism simultaneously prevents cyclic traversals, suppresses repeated links, and ensures that targets referenced by multiple source chunks are included only once in the final context.
	
	\item \textbf{Anchors and broken links management.}  
	Link resolution matches the hyperlink’s source and, when present, its anchor. 
	Links whose source or anchor cannot be mapped to any chunk in the corpus are silently discarded.  
	No placeholder nodes are introduced, and traversal does not backtrack or compensate for unresolved hyperlinks.
	
	\item \textbf{Context-based reranking.}  
	For each valid link, candidate target chunks are reranked using cosine similarity between the link’s local context and the chunk content. Let $l_i$ denote the textual context surrounding a hyperlink and $c_j$ a candidate linked chunk. Each candidate is assigned a score defined as:
	\[
	s(c_j \mid l_i) = \cos\!\left(e(l_i), e(c_j)\right),
	\]
	where $e(\cdot)$ is the embedding function and cosine similarity is computed over $\ell_2$-normalized vectors. Then, for each link context \(l_i\) the candidate chunks are ranked in descending order of $s(c_j \mid l_i)$, and only the \texttt{top\_m} chunks are retained for each expanded link. Reranking is performed at the chunk level; no additional normalization or explicit tie-breaking rules are applied. Query relevance is enforced implicitly through the initial retrieval step that determines the expansion seed set.
	
\end{itemize}

Finally, in both pipelines: \textbf{(v)} the retrieved context and the user query are inserted into their respective placeholders in the prompt template.\footnote{Prompt templates are reported in Appendix \ref{appendix:prompts}.} \textbf{(vi)} The resulting augmented prompt is then passed to the language model, which generates a response conditioned on the retrieved context.



\medskip
Table~\ref{tab:baseline-vs-graph} summarises the main differences between the two approaches across several aspects.

\begin{table}[h]
\centering
\small
\renewcommand{\arraystretch}{1.2}
\begin{tabular}{p{3cm} p{3.5cm} p{8cm}}
\toprule
 & \textbf{RAG} & \textbf{LARAG} \\
\midrule
Document structure  
& Flat chunks  
& Hyperlink‑augmented chunks \\

Metadata  
& Minimal \footnotesize (source, id)  
& Extended \footnotesize (+ anchor, link targets, link-context snippets) \\

Chunking  
& 800 chars + 100 overlap  
& 1000 chars + 150 overlap \\

Embedding model  
& \texttt{text-embedding-3-large}  
& \texttt{text-embedding-3-large} \\

Generation model  
& \texttt{gpt-4o-mini}  
& \texttt{gpt-4o-mini} \\

Vector store  
& Chroma  
& Chroma \\

Retrieval strategy  
& Top-$k$ cosine similarity  
& Cosine similarity + on-the-fly hyperlink expansion \\

Use of links  
& Ignored  
& Parsed and exploited \\

Retrieval granularity  
& Semantic only  
& Semantic + structural \\

Model behaviour  
& Static, context-flat  
& Hyperlink-guided, human-like navigation \\
\bottomrule
\end{tabular}
\caption{Comparison between RAG and LARAG architectures.}
\label{tab:baseline-vs-graph}
\end{table}

\FloatBarrier
\subsection{Prompt design}\label{sec:prompt-design}
Prompt formulation is fundamental in shaping the behaviour of Large Language Models. To evaluate whether different prompt structures interact with the retrieval strategy, we tested four prompt variants that differ in explicitness, structure, and intended level of guidance. These variants range from a minimal zero-shot formulation to prompts that enforce a structured reasoning process or explicitly highlight hyperlinked content.

Table~\ref{tab:prompt-summary} summarises the main characteristics of the four prompt types. All prompts were applied uniformly to both chatbots, with the exception of the hyperlinked prompt. In this case, only LARAG received two separate context blocks, one for the original retrieved content and one for the linked content, while the baseline RAG received a single unified context block. This design allowed us to isolate the effect of explicitly highlighting hyperlink structure during generation. The full text of each prompt is reported in Appendix~\ref{appendix:prompts}.

\begin{table}[h]
\centering
\small
\renewcommand{\arraystretch}{1.15}
\begin{tabular}{p{2.5cm} p{7.5cm}}
\toprule
\textbf{Prompt} & \textbf{Description} \\
\midrule
\textbf{Basic Prompt} & Zero-shot instruction; answer strictly based on the provided context, with minimal guiding structure. \\
\textbf{Role-based} & Assigns the model an expert persona; encourages best practices, warnings, tips, and richer explanations. \\
\textbf{Reasoning} & Requires answering from context but explicitly asks to identify missing information and integrate external knowledge. \\
\textbf{Hyperlinked} & Separates original context from hyperlinked context to highlight connections and reduce linked-content neglect. \\
\bottomrule
\end{tabular}
\caption{Summary of the four prompt templates evaluated. The full text of each prompt is reported in Appendix~\ref{appendix:prompts}.}
\label{tab:prompt-summary}
\end{table}

\section{Experimental results}\label{sec:experimental-results}

A first phase of experimentation was conducted in previous work~\cite{bolognesi2025thesis}, where a benchmark of seven queries, each tested under four prompt formulations, for a total of 28 query–prompt pairs, was used to tune the main parameters of both the baseline RAG and the LARAG architectures. In particular, this experimental phase comprised 252 runs and included a complete evaluation considering cost-related metrics (execution time and token usage), retrieval behaviour (number of retrieved chunks), and semantic quality measured through BERTScore F1.

In the present work, we retain the parameter configuration identified in the previous study and extend the experimental setting in two directions. First, we explicitly compare RAG and LARAG under identical preprocessing configurations, isolating the effect of hyperlink usage by enabling or disabling link-guided traversal. Second, we introduce a larger benchmark of twenty queries (80 query–prompt pairs), refine the quality evaluation by analysing Precision and Recall in addition to F1, and investigate the relationship between semantic quality, context characteristics, and computational cost. This extended evaluation consists of 240 controlled experimental runs.

The following subsections present the benchmark and evaluation metrics, outline the configurations adopted for RAG and LARAG, and report the comparative analysis between the two systems.

\FloatBarrier
\subsection{Experimental setup}

We begin by describing the experimental setup underlying our evaluation, namely the benchmark query set and the metrics used to measure efficiency, retrieval behaviour, and semantic quality.

\subsubsection{Benchmark query set}
To evaluate the performance of the two developed chatbots, a benchmark set of twenty queries was defined. These queries were written by Rulex documentation experts, drawing inspiration from the most frequently asked questions within the user community and from typical information‑seeking patterns observed in real usage. This ensured that the queries were sensible, realistic, and representative of common real‑world use cases. Particular attention was devoted to covering a broad spectrum of Rulex Platform functionalities, ranging from data import to software installation and the configuration of more advanced tasks. The complete list of benchmark queries and the description of the documentation corpus are provided respectively in Appendices~\ref{appendix:queries} and \ref{sec:doc_corpus}.

\FloatBarrier
\subsubsection{Evaluation metrics}
To assess the performance of the two RAG architectures, we rely on four quantitative metrics capturing efficiency, retrieval behaviour, and semantic quality.

\paragraph{Response time.}
Measures the end-to-end latency of the system, from query submission to answer generation. This metric quantifies the computational overhead introduced by hyperlink-aware expansion compared to the baseline system.

\paragraph{Total token count.}
Represents the sum of prompt and output tokens. Since token usage directly affects API cost and latency, this metric allows us to evaluate the trade-off between context size and answer quality.

\paragraph{Retrieved chunks.}
Indicates how much documentation content contributes to the final context. For the baseline RAG this corresponds to the fixed $k$, whereas in the LARAG architecture it varies depending on the parameters controlling hyperlink expansion. This metric reflects the depth and breadth of the retrieved context.

\paragraph{BERTScore.}
We use BERTScore to estimate the semantic similarity between the model output and expert-written gold references. Scores are computed over the same set of references for both systems, enabling a fair comparison of semantic fidelity without requiring human annotation during evaluation.

BERTScore provides three complementary metrics based on semantic embeddings:
\begin{itemize}
	\item \textbf{Precision (P)}: measures how much of the generated answer is semantically supported by the reference.
	\item \textbf{Recall (R)}: measures how much of the reference content is covered by the generated answer.
	\item \textbf{F1}: harmonic mean of P and R, capturing the best overall trade-off between correctness and coverage.
\end{itemize}

\FloatBarrier
\subsection{Optimal configuration selection 
}
\label{subsec:optimal configuration}

This subsection reports the configuration selection process based on an extensive experimental evaluation comprising a total of 252 runs. These experiments were obtained in \cite{bolognesi2025thesis} by systematically combining seven benchmark queries, four prompt formulations, and multiple retrieval configurations for both the baseline RAG system and its link-aware variant, as described in the experimental setup. The resulting metrics form the basis for the configuration choices discussed below.

\paragraph{RAG.}
In the baseline RAG system, the main parameter controlling retrieval is $k$, the number of initially retrieved chunks. In the first set of experiments, $k$ was set to $5$. This choice keeps the prompt length small and focuses on the most relevant chunks, but often led to incomplete answers. Increasing $k$ to $10$ provided broader contextual coverage and significantly improved answer completeness (see Fig. \ref{fig:k5vsk10vsLA} for an example). Therefore, $k=5$ and $k=10$ were retained as reference settings for comparison with LARAG.

\paragraph{LARAG.}
For the link-aware model, the retrieval depth is not controlled solely by $k$, which was fixed to $5$ to enable a fair comparison with the baseline configuration. Instead, hyperlink-guided retrieval is governed by a triple of parameters:
\[
(\texttt{n\_links},\, \texttt{depth},\, \texttt{top\_m}),
\]
with:
\begin{itemize}[leftmargin=10pt]
	\item $\texttt{n\_links}$: the number of outgoing hyperlinks explored from each node;
	\item $\texttt{depth}$: the maximum traversal depth starting from the initially retrieved nodes;
	\item $\texttt{top\_m}$: the maximum number of nodes retained at each expansion step.
\end{itemize}

Although we refer to these steps as a traversal, no explicit graph is ever constructed. Hyperlinks extracted from the HTML documentation are encoded as metadata within each chunk, and traversal is effectively simulated by following these metadata fields during retrieval expansion. This provides a lightweight form of graph navigation: the system exploits author-defined hyperlinks without incurring the overhead of building or maintaining a textual or semantic graph. Several combinations of $(\texttt{n\_links},\, \texttt{depth},\, \texttt{top\_m})$ were evaluated during the preliminary tuning phase, selected to probe different breadth-depth trade-offs in hyperlink expansion while avoiding undesirable behaviours such as uncontrolled growth of candidate subgraphs/paths. Moreover, these configurations were grounded in theoretical considerations about how technical documentation is typically navigated. The optimal configuration was identified through a trade-off between semantic accuracy and computational efficiency, considering F1, average token usage, number of retrieved chunks, and execution time, as shown in Table \ref{tab:linkaware-configurations}.

\begin{table*}[!h]
\centering
\small
\renewcommand{\arraystretch}{1.15}
\sisetup{
  locale = US,            
  detect-weight = true,
  detect-family = true
}
\begin{tabular}{
  l
  S[table-format=2.0]
  S[table-format=4.0]
  S[table-format=2.3]
  S[table-format=1.4]
}
\toprule
\textbf{Configuration} & \textbf{Chunks} & \textbf{Tokens} & \textbf{Time (s)} & \textbf{F1} \\
\midrule
RAG (k = 5) & \textcolor{blue}{\textbf{5}} & \textcolor{blue}{\textbf{1225}} & 8.52 & \textcolor{red}{\textbf{0.8227}}\\
RAG (k = 10) & 10 & 2035 & 10.65 & \textcolor{red}{\textbf{0.8229}} \\
LARAG (0,0,0) & \textcolor{blue}{\textbf{5}} & 1285 & \textcolor{blue}{\textbf{8.23}} & \textcolor{red}{\textbf{0.8227}} \\
LARAG (1,1,1) & \textbf{8} & \textbf{1669} & \textbf{9.72} & 0.8253 \\
LARAG (1,3,1) & 10 & 2047 & 13.78 & \textcolor{blue}{\textbf{0.8276}} \\
LARAG (1,3,3) & \textcolor{red}{\textbf{41}} & \textcolor{red}{\textbf{4387}} & 15.54 & \textcolor{red}{\textbf{0.8007}} \\
LARAG (2,1,2) & 12 & 2213 & 10.89 & 0.8252 \\
LARAG (3,1,3) & 15 & 2607 & 12.02 & 0.8254 \\
LARAG (3,2,3) & \textcolor{red}{\textbf{40}} & \textcolor{red}{\textbf{3674}} & 15.71 & \textcolor{red}{\textbf{0.7973}} \\
\bottomrule
\end{tabular}
\caption{Summary of performance metrics for each configuration of RAG with \(k\) chunks and LARAG$(\texttt{n\_links},\, \texttt{depth},\, \texttt{top\_m})$ with fixed \(5\) chunks. Note that LARAG (1,1,1) is equivalent to the baseline RAG system with \(k=5\).}
\vspace{2pt}
{\footnotesize
	\textit{Legend.} Values in \textcolor{blue}{\textbf{bold blue}} indicate the best scores for each metric, whereas values in \textcolor{red}{\textbf{bold red}} denote the worst ones. \textbf{Bold, non-coloured} values highlight the metrics in which configuration $(1,1,1)$ outperforms configuration $(1,3,1)$.
}
\label{tab:linkaware-configurations}
\end{table*}

This analysis revealed that configuration $(1,1,1)$ offered the most favourable balance across these criteria. As reported in Table~\ref{tab:linkaware-configurations}, although it does not achieve the highest F1, result obtained by configuration $(1,3,1)$, it attains one of the \emph{highest} F1 while requiring substantially fewer chunks, fewer generated tokens, and a noticeably lower response time. This more efficient quality-cost profile makes $(1,1,1)$ particularly suitable for real-time or resource-constrained deployments, and it was therefore selected as the final configuration for the comparative evaluation.

\medskip
We also report in Table \ref{tab:aggGio} the aggregate comparative results originally obtained in~\cite{bolognesi2025thesis}, where the baseline RAG configurations (\(k=5\) and \(k=10\)) were compared against LARAG \((1,1,1)\)

\begin{table*}[!h]
	\centering
	\small
	\begin{tabular}{l c c c c}
		\toprule
		\textbf{Configuration} & \textbf{Chunks} & \textbf{Tokens} & \textbf{Time (s)} & \textbf{F1} \\
		\midrule
		RAG (k = 5) & 5 & 1226 & 8.52 & 0.8227 \\
		RAG (k = 10) & 10 & 2035 & 10.65 & 0.8229\\
		LARAG (1,1,1) & 8 & 1669 & 9.72 & 0.8253\\
		\bottomrule
	\end{tabular}
	\caption{Aggregate results from \cite{bolognesi2025thesis}.}
	\label{tab:aggGio}
\end{table*}

While these results already suggested the effectiveness of hyperlink-aware retrieval, the comparison did not isolate the contribution of hyperlink traversal from other architectural and preprocessing differences. The present work builds directly on this evidence by re-evaluating RAG and LARAG under identical preprocessing conditions, explicitly isolating the effect of hyperlink usage.




\FloatBarrier
\subsection{Comparative evaluation: RAG vs.\ LARAG}

This comparative evaluation is based on a total of 240 experiments, obtained by combining 20 benchmark queries, four prompt formulations, and three retrieval configurations selected for comparison: a RAG configuration with retrieval depth \(k=5\), a RAG configuration with \(k=10\), and LARAG with graph‑traversal parameters \((1,1,1)\) and \(k=5\). However, differently from the experimental setting adopted in \cite{bolognesi2025thesis}, our goal here is to isolate the effect of hyperlink usage. To this end, the two RAG variants are implemented as instances of LARAG with graph‑traversal explicitly disabled, i.e., with traversal parameters set to \((0,0,0)\), while preserving the same preprocessing pipeline. This setting ensures that any observed performance differences cannot be attributed to variations in chunking strategy, chunk size, overlap, HTML parsing, or metadata construction, but solely to the activation or deactivation of hyperlink‑guided retrieval.

In the following, we denote these three models as:


\begin{center}
	\textbf{RAG\_k5} {(\footnotesize LARAG\((0,0,0)\), \(k=5\))}, \textbf{RAG\_k10} {\footnotesize (LARAG\((0,0,0)\), \(k=10\))}, \textbf{LARAG} {\footnotesize (LARAG\((1,1,1)\), \(k=5\))}.
\end{center}


\begin{figure}[!htbp]
	\centering
	\includegraphics[width=0.9\textwidth]{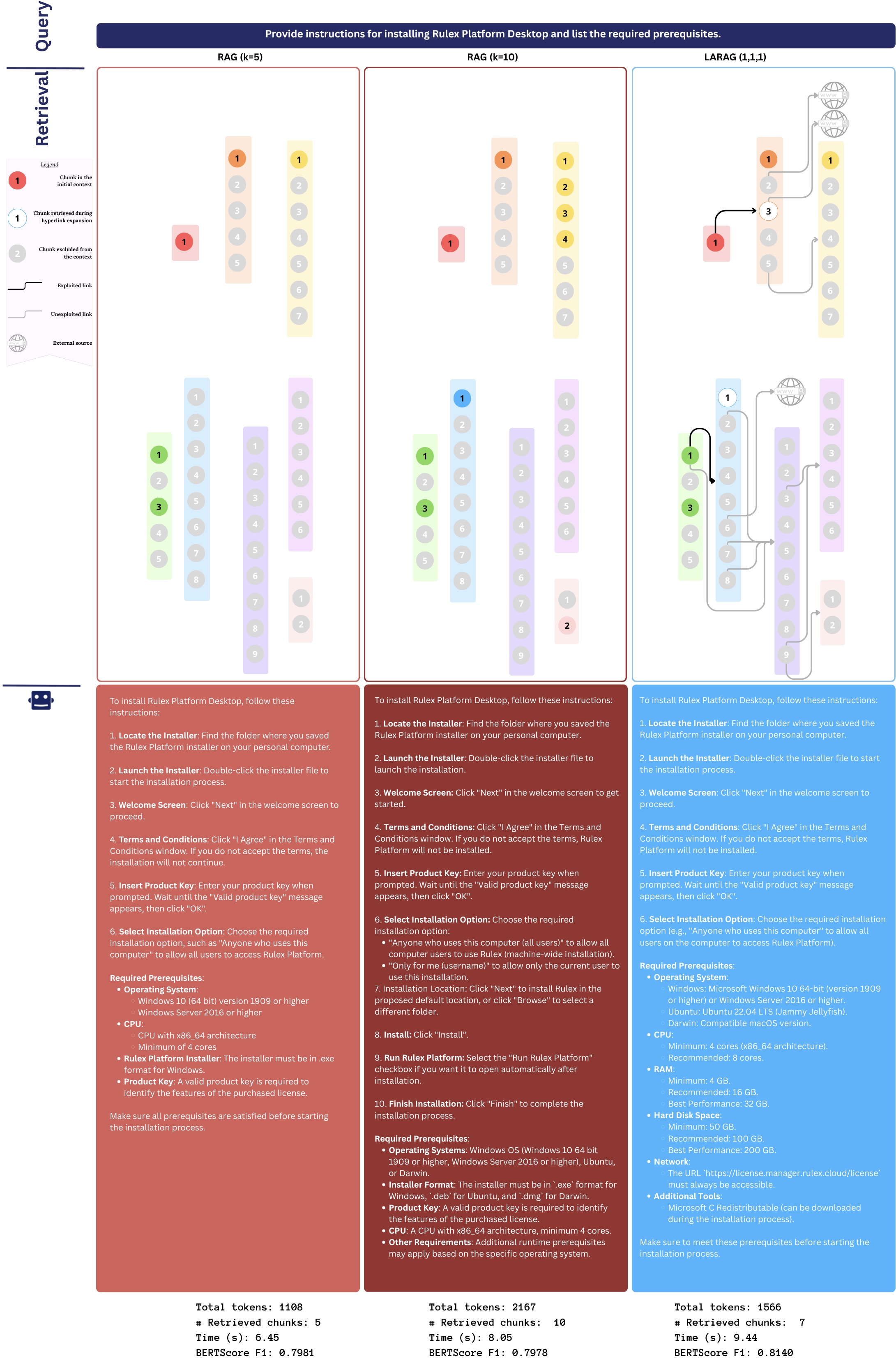}
	\caption{\footnotesize Comparison between the responses generated using RAG\_k5 (left), RAG\_k10 (center), and LARAG (right). The selected query is Query~3 (see Appendix~\ref{appendix:queries}) and the prompt template is the Basic one (see Appendix~\ref{appendix:prompts}). The comparison shows that increasing \(k\) improves answer completeness only marginally, whereas LARAG retrieves more relevant topics by following semantically related chunks reached through hyperlink expansion (e.g., it captures all the required prerequisites, which are only partially covered when increasing \(k\)). From a cost perspective, RAG\_k5 is the most efficient but under‑informative, RAG\_k10 incurs substantially higher token usage and retrieval volume, while LARAG achieves better coverage with fewer retrieved chunks and significantly fewer tokens; execution time is slightly higher but remains comparable, as discussed in the subsequent analysis. The  full textual content of the retrieved chunks, is provided in Appendix~\ref{appendix:example}.}
	\label{fig:k5vsk10vsLA}
\end{figure}

The analysis first examines their cost and answer‑quality behaviour, and then investigates how answer length and prompting strategies influence their performance.

\paragraph{Cost and answer quality analysis.}

\begin{table*}[h!]
	\centering
	\small
	\renewcommand{\arraystretch}{1.2}
	\begin{tabular}{p{2.5cm} c c c c c c}
		\toprule
		\textbf{Configuration} & \textbf{Total tokens} & \textbf{Retrieved chunks} & \textbf{Time (s)} & \textbf{P} & \textbf{R} & \textbf{F1} \\
		\midrule
		
		RAG\_k5 & \textcolor{blue}{\textbf{1135}} & \textcolor{blue}{\textbf{5}} & \textcolor{blue}{\textbf{8.22}} & 0.8295 & \textcolor{red}{\textbf{0.8430}} & \textcolor{red}{\textbf{0.8360}} \\
		RAG\_k10 & \textcolor{red}{\textbf{1853}} & \textcolor{red}{\textbf{10}} & 8.78 & \textcolor{red}{\textbf{0.8290}} & \textcolor{blue}{\textbf{0.8471}} & 0.8378 \\
		LARAG & \textbf{1527} & \textbf{8} & 11.37   & \textcolor{blue}{\textbf{0.8303}} & \textbf{0.8469} & \textcolor{blue}{\textbf{0.8382}} \\
		\bottomrule
	\end{tabular}
	\caption{Aggregate results of our analysis.}
	\vspace{2pt}
	{\footnotesize
		\textit{Legend.} Values in \textcolor{blue}{\textbf{blue}} indicate the best scores for each metric, whereas values in \textcolor{red}{\textbf{red}} denote the worst ones. \textbf{Bold, non-coloured} values highlight the metrics in which configuration LARAG outperforms or is comparable to RAG\_k10.
	}
	\label{tab:agg}
\end{table*}

Table~\ref{tab:agg} summarises the overall cost and quality of the three RAG configurations.

\begin{itemize}
	\item RAG\_k5 is the most economical setting: it consistently retrieves 5 chunks, produces the lowest token count (1135 on average), and achieves the shortest latency (8.22\,s). This efficiency, however, comes with reduced completeness, as reflected by the lowest F1 and by several incomplete answers (e.g., Fig.~\ref{fig:k5vsk10vsLA}).
	\item RAG\_k10 retrieves the largest amount of context (10 chunks, 1853 tokens). While recall marginally improves, the gains in quality are limited with respect to the substantial increase in cost.
	\item LARAG retrieves fewer chunks than RAG\_k10 (about 8 on average, 1527 tokens) yet attains the highest F1. This suggests that the hyperlink-guided expansion improves relevance rather than merely increasing retrieval volume. Its higher mean latency (11.37\,s) should be interpreted with caution: the experiments were executed locally, where timing is sensitive to system variability, and previous evaluations in~\cite{bolognesi2025thesis} showed that LARAG can match or even surpass RAG\_k10 in speed, as reported in Table \ref{tab:aggGio}. Consistently with this, the per-query execution-time plots in Appendix~\ref{sec:per_query_perf} show that LARAG is often comparable to RAG\_k10, with only small query-level fluctuations rather than a systematic slowdown. A detailed explanation of the per-query variability in execution time, and of why latency does not always align with token usage, is provided in Appendix~\ref{appendix:aggr_results}. Moreover, from a user‑experience perspective, a slightly longer latency is generally preferable if it yields a more useful and contextually complete answer rather than a faster but less informative one.
\end{itemize}

\begin{figure}[!htbp]
	\centering
	\includegraphics[width=0.85\textwidth]{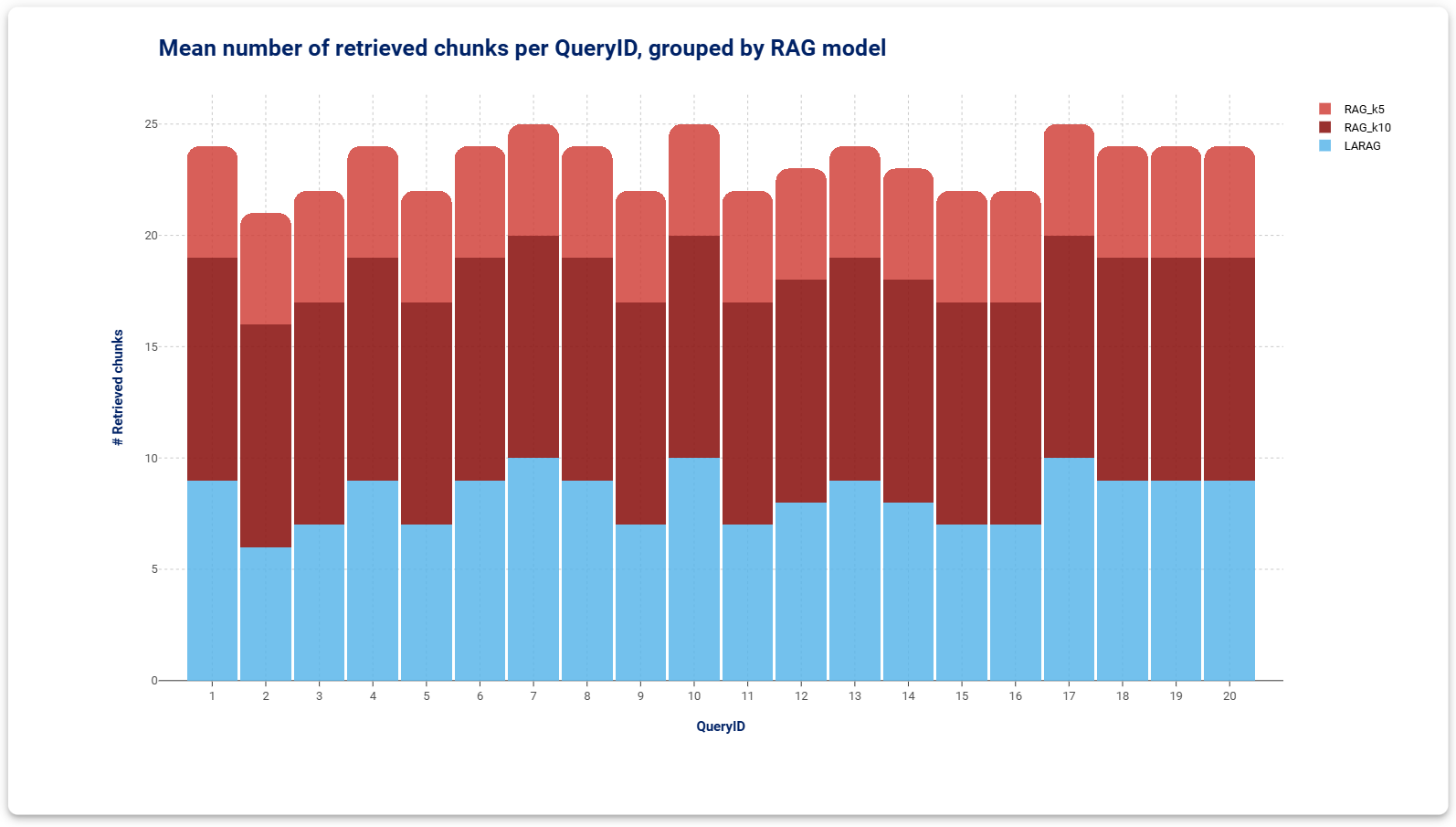}
	\includegraphics[width=0.85\textwidth]{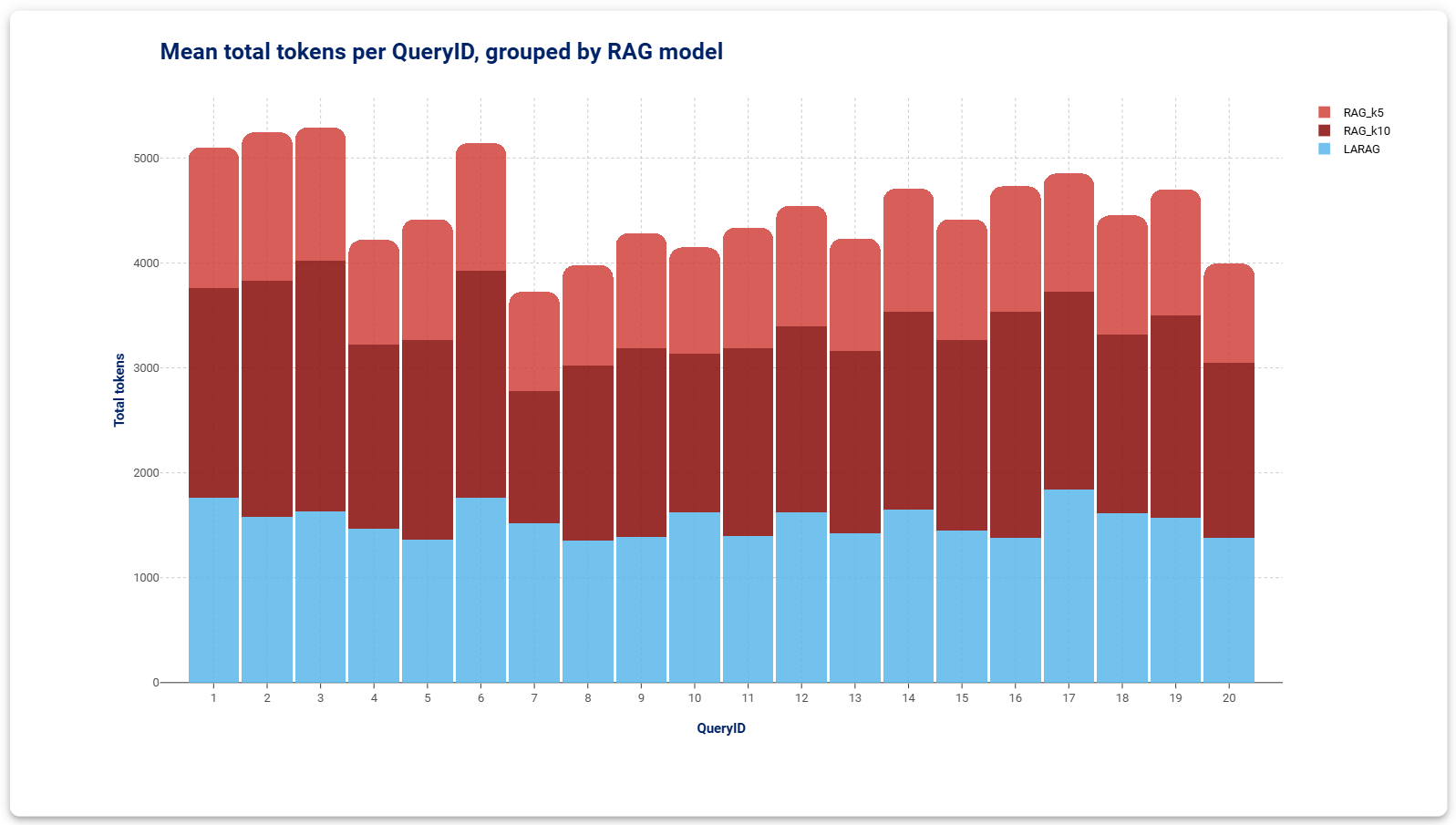}
	\includegraphics[width=0.85\textwidth]{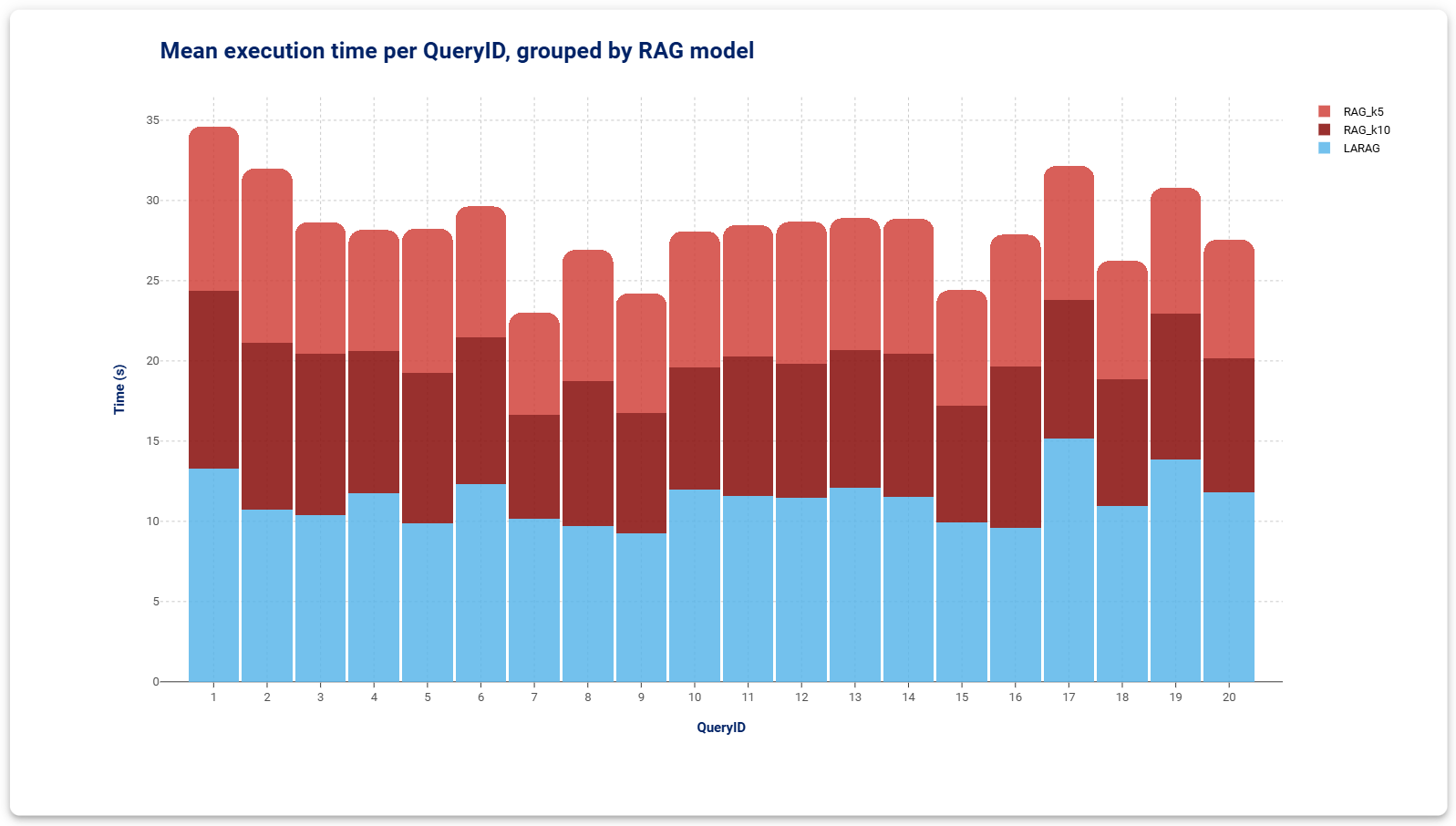}
	\caption{Average retrieved chunks, total tokens and execution time aggregated by model. Images generated with Rulex Studio.}
	\label{fig:qr_chunks_tokens_exec_all}
\end{figure}

These patterns appear also in the per-query distributions (Fig~\ref{fig:qr_chunks_tokens_exec_all}).

%


Overall, RAG\_k5 offers minimal cost, RAG\_k10 trades efficiency for coverage, and LARAG provides the best balance between cost and answer quality.
The full per-query and per-prompt breakdowns underlying these aggregate trends are reported in Appendix~\ref{appendix:details}.

\paragraph{Cost-quality correlation.}\label{subsec:cost-quality}
To further characterise the cost-quality trade-off, we also examined the correlation between F1 and two cost-related metrics: total token count and execution time. Results are reported in Table~\ref{tab:corr_cost_quality}. As one could expect, all models exhibit a negative correlation, but notably, LARAG shows the weakest correlation with total tokens, suggesting that its hyperlink-aware expansion retrieves more informative context without relying on large increases in input size.

\begin{table}[h!]
	\centering
	\small
	\begin{tabular}{lcc}
		\toprule
		\textbf{Configuration} & \textbf{corr(F1, total tokens)} & \textbf{corr(F1, exec.\ time)} \\
		\midrule
		RAG\_k5  & \textcolor{red}{\textbf{-0.430}} & \textcolor{red}{\textbf{-0.311}} \\
		RAG\_k10 & -0.378 & \textcolor{blue}{\textbf{-0.270}} \\
		LARAG   & \textcolor{blue}{\textbf{-0.280}} & \textbf{-0.289} \\
		\bottomrule
	\end{tabular}
	\caption{Correlation between F1 and cost-related metrics (lower magnitude is better).}
	{\footnotesize
		\textit{Legend.} Values in \textcolor{blue}{\textbf{bold blue}} indicate the best scores for each metric, whereas values in \textcolor{red}{\textbf{bold red}} denote the worst ones. The \textbf{bold, non-coloured} value highlights a value for LARAG that is comparable with the corresponding value for RAG\_k10.
	}
	\label{tab:corr_cost_quality}
\end{table}

Fine-grained results for each QueryID are reported in Appendix~\ref{sec:per_query_perf}, which confirm that these aggregate trends hold consistently across the full benchmark.

\paragraph{Length effects.}\label{subsec:length_effects}

Lengths \texttt{len\_ref}, \texttt{len\_pred} are computed as word counts, as token-level information is not available for either the reference or the generated response (only the total number of tokens used during generation is provided). Table~\ref{tab:length_corr} reports the correlation between F1 and reference or prediction length for each model. In all cases, F1 decreases as length increases, with reference length having the stronger effect. LARAG is the most robust to this degradation, while RAG\_k5 is least affected by prediction length.




\begin{table}[h!]
	\centering
	\small
	\begin{tabular}{l c c c}
		\toprule
		\textbf{Correlations} & \textbf{RAG\_k5} & \textbf{RAG\_k10} & \textbf{LARAG}\\
		\midrule
		corr(F1, \texttt{len\_ref}) & -0.551 & -0.554 & \textbf{-0.503} \\
		corr(F1, \texttt{len\_pred}) & \textbf{-0.323} & -0.344 & -0.358 \\
		\bottomrule
	\end{tabular}
	\caption{Correlation between F1 and reference/answer length.}
	\label{tab:length_corr}
\end{table}

The stronger negative correlation between reference length and F1 (Table~\ref{tab:length_corr}) is further confirmed by the full correlation matrix reported in Appendix~\ref{sec:corr_length_BERTScore}.


To analyse performance across different length regimes, reference lengths are grouped into four quantile-based bins using \texttt{qcut}. Table~\ref{tab:length_bins} shows the mean F1 per bin. A clear regime split emerges: RAG\_k5 performs best on short references, RAG\_k10 leads on medium-length ones, and LARAG achieves the highest scores for long references.

\begin{table}[h!]
	\centering
	\small
	\begin{tabular}{l c c c}
		\toprule
		\textbf{\texttt{len\_ref} bins} & \textbf{RAG\_k5} & \textbf{RAG\_k10} & \textbf{LARAG} \\
		\midrule
		(53.999, 234.75]  & \textbf{0.8494} & 0.8483 & 0.8483 \\
		(234.75, 322.0]   & 0.8365 & 0.8397 & \textbf{0.8405} \\
		(322.0, 575.5]    & 0.8405 & \textbf{0.8440} & 0.8412 \\
		(575.5, 1158.0]   & 0.8175 & 0.8192 & \textbf{0.8228} \\
		\bottomrule
	\end{tabular}
	\caption{Mean F1 by reference-length bins.}
	\label{tab:length_bins}
\end{table}



\paragraph{Prompting strategies effects.}\label{subsec:prompting}

\begin{table}[h!]
	\centering
	\small
	\begin{tabular}{l c c c c p{5cm}}
		\toprule
		\textbf{Prompt} & \textbf{Clarity} & \textbf{Completeness} & \textbf{Ideal User} & \textbf{Verbosity} & \textbf{Typical Behaviour} \\
		\midrule
		\textbf{Basic} & High & Medium & Expert & Low &
		Concise, step-oriented answers; minimal contextualisation; good precision. \\
		\textbf{Role-based} & High & High & Beginner & High &
		Pedagogical tone; includes tips/warnings; richer (often off-reference content). \\
		\textbf{Reasoning} & Low & Low & Expert & Low &
		Analytical but cautious; often states insufficient context; weakest semantic alignment. \\
		\textbf{Hyperlinked} & High & High & Beginner & Medium &
		Structured, well-organised responses leveraging internal links; balanced completeness and clarity. \\
		\bottomrule
	\end{tabular}
	\caption{Qualitative comparison of answer characteristics across prompting strategies.}
	\label{tab:prompt_qualitative}
\end{table}

Prompting has a measurable but secondary effect on semantic accuracy and a substantial impact on verbosity and structure. Execution-time plots (Fig.~\ref{fig:exec_by_model} in Appendix \ref{appendix:details}) show consistent trends: Role-based prompts generate the largest token budgets and highest latencies, while Basic prompts remain the most efficient.

In terms of BERTScore, differences across prompting strategies are relatively small: Basic prompts achieve the highest macro-F1, whereas Role-based prompts obtain slightly lower scores, partly due to the inclusion of additional explanatory content that is not always reflected in the reference answers.
Since similarity-based metrics such as BERTScore primarily capture semantic overlap with a gold reference, they may only partially reflect responses that include additional but potentially useful information. For this reason, quantitative evaluation is complemented by a qualitative analysis, summarised in Table~\ref{tab:prompt_qualitative}, which highlights differences in clarity, completeness, and instructional style that are not captured by automatic metrics alone.

Appendix~\ref{sec:per_prompt_perf} provides per-prompt tables and plots that further illustrate the qualitative differences described above.

\paragraph{Summary.}
The three configurations exhibit distinct and consistent profiles.

\begin{itemize}
	\item \textbf{RAG\_k5} offers the smallest computational load and often achieves the highest Precision, as it retrieves fewer chunks and therefore introduces less additional content; however, this limits coverage and results in systematically lower Recall and F1.
	\item \textbf{RAG\_k10} leverages broader context and consistently maximises Recall, but the increased retrieval volume does not translate into the best overall accuracy.
	\item \textbf{LARAG} provides the most favourable quality--cost trade-off: it retrieves fewer chunks than RAG\_k10 while achieving the highest macro-F1 and competitive Precision, and it remains more robust to reference-length variation.
\end{itemize}

Figure~\ref{fig:k5vsk10vsLA} provides a concrete illustration of these differences on a representative query. RAG\_k10 produces the most verbose output, reflecting the larger retrieved context, while RAG\_k5 yields a more compact answer at the cost of reduced coverage. By contrast, LARAG achieves a well-structured and comprehensive response with a substantially smaller token budget, highlighting how hyperlink-aware retrieval can improve answer quality by selecting more relevant context rather than increasing retrieval volume. Prompting strategies primarily affect verbosity and perceived clarity rather than semantic accuracy: Basic prompts maximise efficiency, Role-based and Hyperlinked prompts improve readability at higher computational cost, and Reasoning prompts underperform across metrics (see Appendix~\ref{sec:per_prompt_perf}). Overall, hyperlink-aware retrieval improves answer quality by selecting more relevant context rather than increasing retrieval volume, making LARAG the most balanced and effective configuration across heterogeneous query types.

\section{Conclusions and future work}\label{sec:conclusions}

This work investigated whether enriching retrieval with hyperlink structure can improve the effectiveness and efficiency of RAG systems applied to technical documentation. Using Rulex Platform as a case study, we compared a standard RAG configuration with a link‑aware variant, LARAG, that expands the retrieved context by following documentation hyperlinks. The evaluation, conducted on a benchmark of twenty diverse queries and four prompting strategies, shows that hyperlink‑aware retrieval provides a more favourable quality–cost trade-off than relying solely on semantic similarity.

Across all metrics, LARAG achieves the highest macro-level F1 while retrieving fewer chunks than RAG\_k10. Precision also improves, suggesting that hyperlink expansion selects more relevant evidence rather than simply increasing retrieval volume. At the same time, the cost analysis demonstrates that link-aware retrieval remains computationally lightweight, with token usage clearly below that of RAG\_k10 and execution times comparable once query-level variation and system-level noise are taken into consideration. Length‑sensitivity analysis further indicates that the link-aware approach is more robust on long or multi‑step queries, where contextual coherence matters most.

Prompting strategies modulate verbosity and user‑perceived clarity more strongly than semantic accuracy: Basic prompts remain the most efficient, while Role-based and Hyperlinked prompts favour completeness and structure at higher cost. By exploiting relations that reflect the navigational intent of documentation authors, the proposed approach improves grounding while remaining lightweight and robust, making it well suited for scalable AI assistants in complex technical domains. 

Overall, these findings show that hyperlink-aware retrieval can effectively improve RAG performance by prioritising semantically connected content rather than expanding context indiscriminately. This outcome highlights the value of leveraging documentation structure, beyond purely embedding-based similarity, when designing scalable and reliable AI assistants for complex technical domains. More broadly, LARAG can be understood as a retrieval strategy that simulates local graph exploration at query time, while deliberately avoiding the construction, maintenance, or querying of an explicit graph structure. This choice isolates the contribution of author-defined structural signals without introducing the complexity of graph indexing or graph-native retrieval models.


\medskip

Building on these results, several opportunities emerge for extending and enhancing structure‑aware retrieval in future work.


A first direction concerns the type of graph structure exploited during retrieval. While our approach leverages the explicit hyperlink topology of the corpus, it does not incorporate the induced semantic or textual graphs explored in recent Graph-RAG systems \cite{hu-etal-2025-grag,knollmeyer2025document,dong2024don,han2024retrieval}. Hyperlinks capture the navigational and organizational intent of the authors, whereas induced graphs model latent conceptual relations that are not explicitly encoded in the documentation. These two sources of structure are naturally complementary, suggesting the potential benefits of retrieval strategies that integrate hypertextual and induced semantic edges within a unified framework.


Further gains may arise from the embedding models used for retrieval. Because chunk selection depends directly on embedding quality, alternative architectures, domain-adapted embeddings, and fine‑tuning strategies represent promising directions \cite{incitti2023beyond,fatemi2023talk,tan2023walklm,kraivsnikovic2025fine}; current work already explores these possibilities.

The traversal strategy itself also offers room for extension. While our depth-first hyperlink expansion improves over top-$k$ retrieval, it does not yet exploit the full expressive power of the documentation graph. Dynamic edge weighting and query‑aware traversal strategies \cite{lau2026breaking,zhouquery}, centrality‑ or community‑aware heuristics \cite{rajeh2023comparative, ismaeel2025comparing}, hybrid breadth–depth exploration \cite{cosson2023efficient}, and reinforcement‑based traversal policies \cite{lin2025comprehensive, li2023reinforcement} represent promising classes of techniques that could be adapted to hyperlink‑level retrieval. Integrating even a subset of these strategies may enable more adaptive exploration tailored to heterogeneous query types.

Finally, safety and reliability considerations merit attention \cite{Huang_2025}. Even with improved grounding, structure‑aware retrieval may surface outdated or deprecated material when legacy content remains reachable through historical or cross‑referenced hyperlinks. Ensuring version‑awareness, tracking deprecations, and filtering stale content will be important for reliable deployment, as recent analyses highlight how graph‑guided retrieval can drift toward structurally prominent but semantically outdated regions \cite{lau2026breaking} and how inconsistent or obsolete knowledge sources contribute to misinformation risks in LLMs \cite{Huang_2025}.

\section*{Acknowledgments}
This work was carried out within the framework of the project \textit{Programma Regionale Fondo Sociale Europeo+ 2021--2027, Priorità~2 -- Istruzione e Formazione -- ESO~4.6 (OS-f)}, in collaboration with and with the contribution of \textit{Rulex Innovation Labs S.r.l.}.

\bibliographystyle{acm}
\bibliography{refs}

\newpage
\appendix
\section{Prompt templates}
\label{appendix:prompts}

This appendix reports the full text of the four prompt templates used in the evaluation.  
Each template corresponds to one of the prompting strategies discussed in Section~\ref{sec:prompt-design} and is presented exactly as implemented during experimentation.

\FloatBarrier
\subsubsection*{Prompt 1 — Basic prompt}

\begin{promptbox}
	\begin{verbatim}
		Answer the question based only on the following context:
		
		{context}
		
		---
		
		Answer the question based on the above context: {question}
	\end{verbatim}
\end{promptbox}

\paragraph{Description.}
A minimal zero-shot prompt designed to constrain the model to the provided context only.  
It explicitly separates the context from the question, reducing hallucinations and improving interpretability.

\bigskip

\FloatBarrier
\subsubsection*{Prompt 2 — Role-based prompt}
\begin{promptbox}
	\begin{verbatim}
		You are a technical assistant specializing in Rulex documentation.
		Answer the question using best practices, potential problems,
		and expert recommendations. 
		If applicable, include a "Warning" or "Tip" section.
		
		CONTEXT: {context}
		QUESTION: {question}
		ANSWER:
	\end{verbatim}
\end{promptbox}

\paragraph{Description.}
Assigns the model an expert persona and encourages richer, more human-like responses including recommendations, warnings, and tips.  
The structured CONTEXT–QUESTION–ANSWER format reduces ambiguity and improves answer reliability.

\bigskip

\FloatBarrier
\subsubsection*{Prompt 3 — Reasoning prompt}
\begin{promptbox}
	\begin{verbatim}
		Answer the question based only on the following context. 
		If the context does not provide sufficient information, explicitly
		state which details are missing and supplement them with 
		external documentation.
		
		CONTEXT: {context}
		QUESTION: {question}
		ANSWER:
	\end{verbatim}
\end{promptbox}

\paragraph{Description.}
Combines strict context grounding with explicit handling of missing information.  
The model is instructed to identify gaps and integrate external knowledge only when necessary, improving completeness and transparency.

\bigskip

\FloatBarrier
\subsubsection*{Prompt 4 (for LARAG) — Hyperlinked prompt}
\begin{promptbox}
	\begin{verbatim}
		Original context:
		{original_context}
		
		---
		
		Additional context (linked):
		{linked_context}
		
		---
		
		Question:
		{question}
		
		Please use both sections of context to answer the question 
		comprehensively. Carefully consider the information from 
		both the original context and the linked context.
	\end{verbatim}
\end{promptbox}

\paragraph{Description.}
This prompt explicitly separates the original retrieved content from the additional hyperlinked content. The goal is to force the model to process both sources rather than implicitly ignoring the linked content, a behaviour common when context blocks are merely concatenated.

\FloatBarrier
\subsubsection*{Prompt 4 (for RAG) — Unified context prompt}
\begin{promptbox}
	\begin{verbatim}
		CONTEXT:
		{context}
		
		QUESTION:
		{question}
		
		Please use the above context to answer the question comprehensively.
	\end{verbatim}
\end{promptbox}

\paragraph{Description.}
This version is used for the baseline chatbot, which does not receive any content retrieved through hyperlinks. It presents the entire retrieved context in a single block, mirroring the structure of traditional RAG prompting. This allows a direct comparison with the Hyperlinked Prompt, isolating the effect of explicit hyperlink segmentation.

\newpage
\section{Benchmark and documentation resources}

This appendix presents the two resources underlying our evaluation: the documentation corpus that serves as the knowledge base for both RAG and LARAG, and the complete benchmark query set used to assess retrieval and generation quality.

\FloatBarrier
\subsection{Documentation corpus: Rulex Platform}\label{sec:doc_corpus}
\begin{figure}[!htbp]
	\centering
	\begin{subfigure}[t]{0.48\linewidth}
		\centering
		\includegraphics[width=\linewidth]{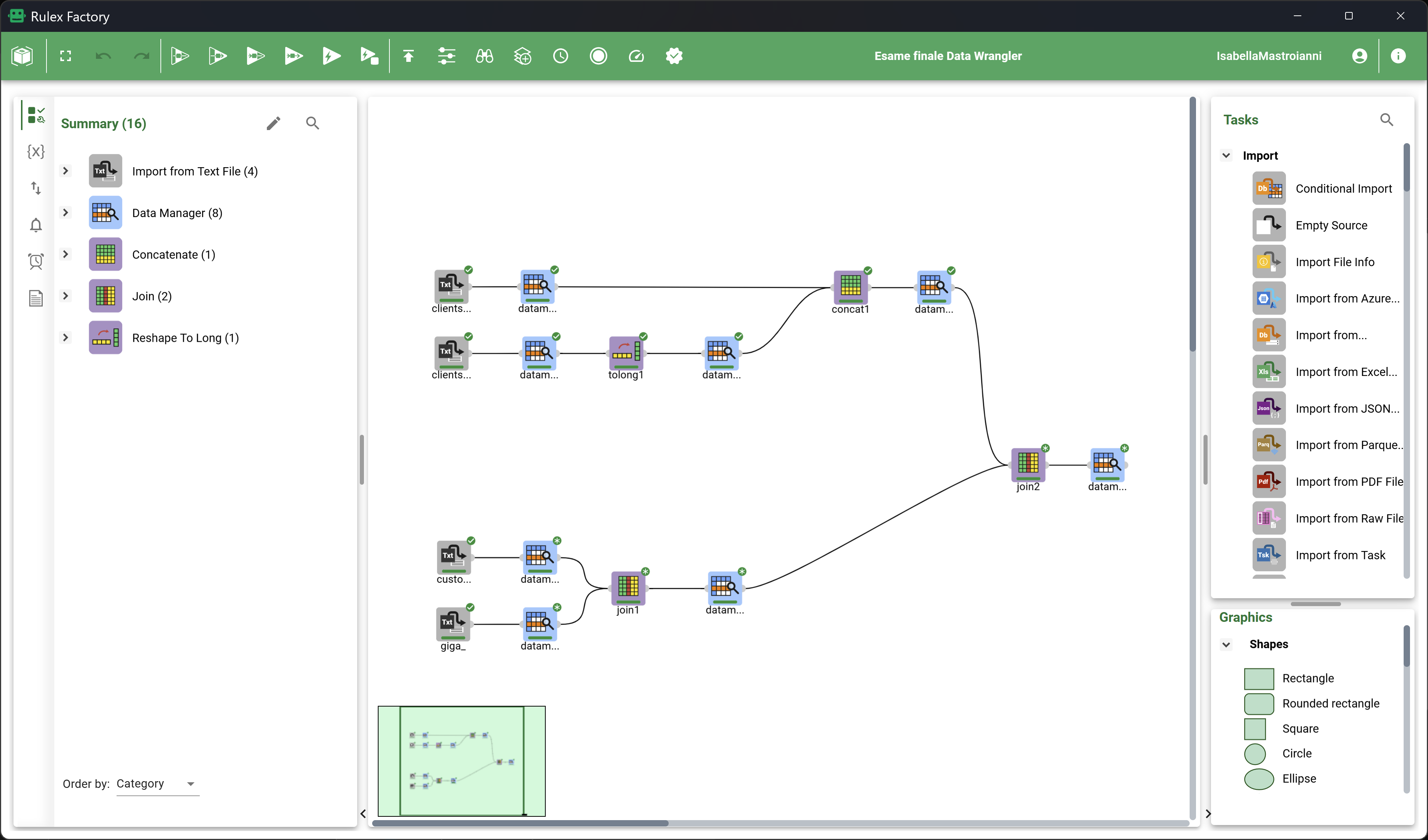}
		\caption{ }
		\label{fig:Factory dashboard}
	\end{subfigure}
	\hfill
	\begin{subfigure}[t]{0.48\linewidth}
		\centering
		\includegraphics[width=\linewidth]{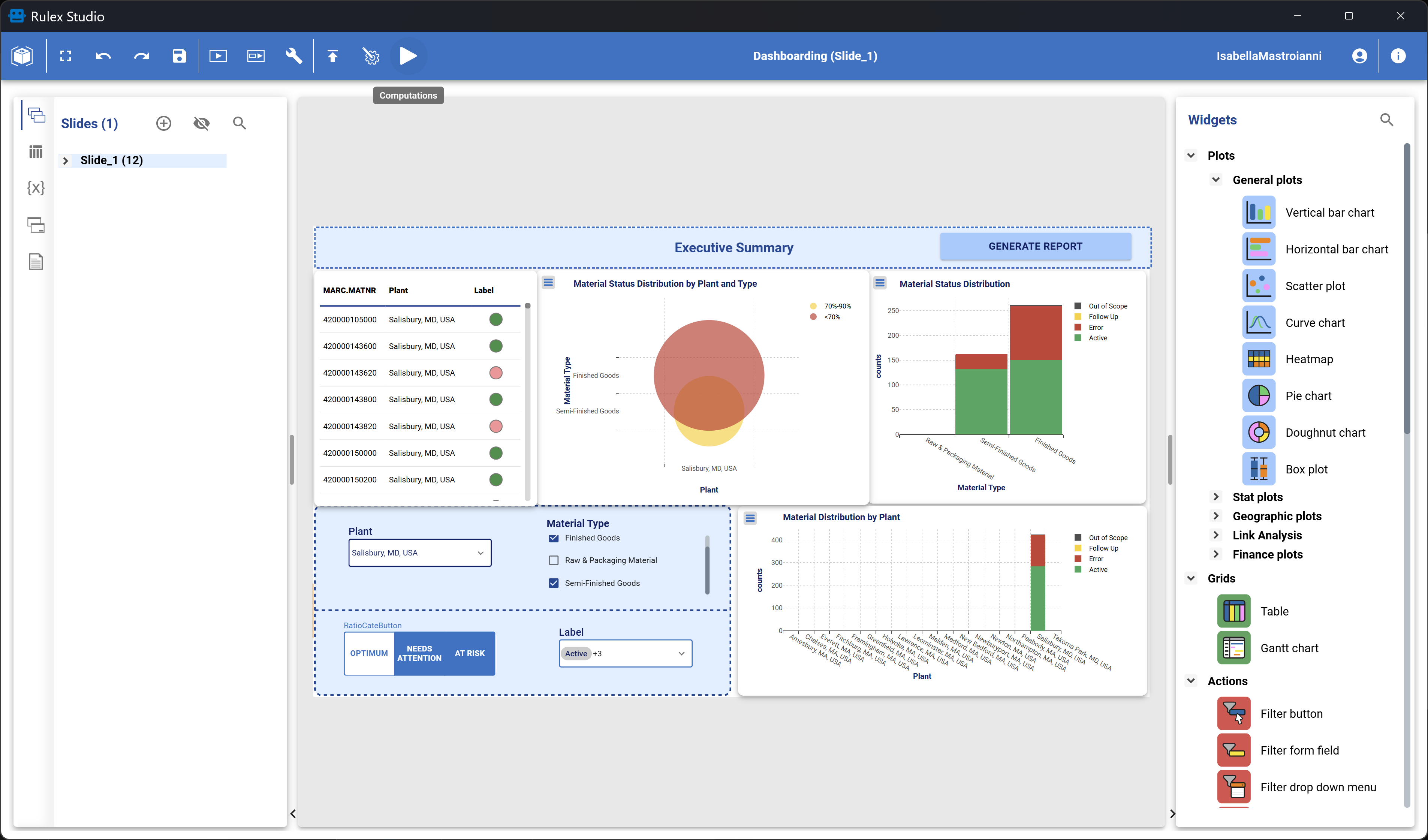}
		\caption{ }
		\label{fig:Studio dashboard}
	\end{subfigure}
	\caption{(a) dashboard of Rulex Factory; (b) dashboard of Rulex Studio.}
	\label{fig:example Rulex dashboards}
\end{figure}

The documentation used in this work as the corpus for both the RAG chatbot and the enhanced LARAG chatbot is the Rulex Platform Documentation, version 1.4.x, consisting of 239 files \cite{rulex_docs_14x}. Rulex Platform is a data management and decision intelligence platform that enables users to build, monitor, integrate, run, and maintain enterprise-level solutions. It consists of two core, fully integrated components: Rulex Factory and Rulex Studio. Rulex Factory is the processing engine that allows users to build data flows and monitor results, offering tools for data import, transformation, optimization, rule management, and scheduling (Fig. \ref{fig:Factory dashboard}); Rulex Studio is the front-end environment used to create interactive dashboards for data visualization (Fig. \ref{fig:Studio dashboard}).

This documentation corpus was chosen because it represents a real-world use case of technical documentation and presents a rich hypertext structure characterized by many internal links and cross-references between pages, making it a valuable framework for assessing the effectiveness of hyperlink-aware retrieval within a RAG system. In particular, the content is organized into three main sections:
\begin{itemize}
	\item \textit{Rulex Platform}, which provides an overview of the platform as a data management system and introduces its two main components, Factory and Studio
	;
	\item \textit{Rulex Factory}, which documents the components and functionalities of the processing engine, describing how users can design and manage data flows through an interactive canvas. This section explains tasks for data import, transformation, optimization, rule management, and scheduling, along with their configuration options and monitoring capabilities;
	\item \textit{Rulex Studio}, which details the front-end environment for data visualization and user interaction. It explains how dashboards and Views are built using drag-and-drop widgets, how real-time filters and data editing are configured, and how dashboards are synchronized with Rulex Factory and exported as reports in PDF format.
\end{itemize}

\begin{figure}[!htbp]
	\centering
	\includegraphics[width=0.75\textwidth]{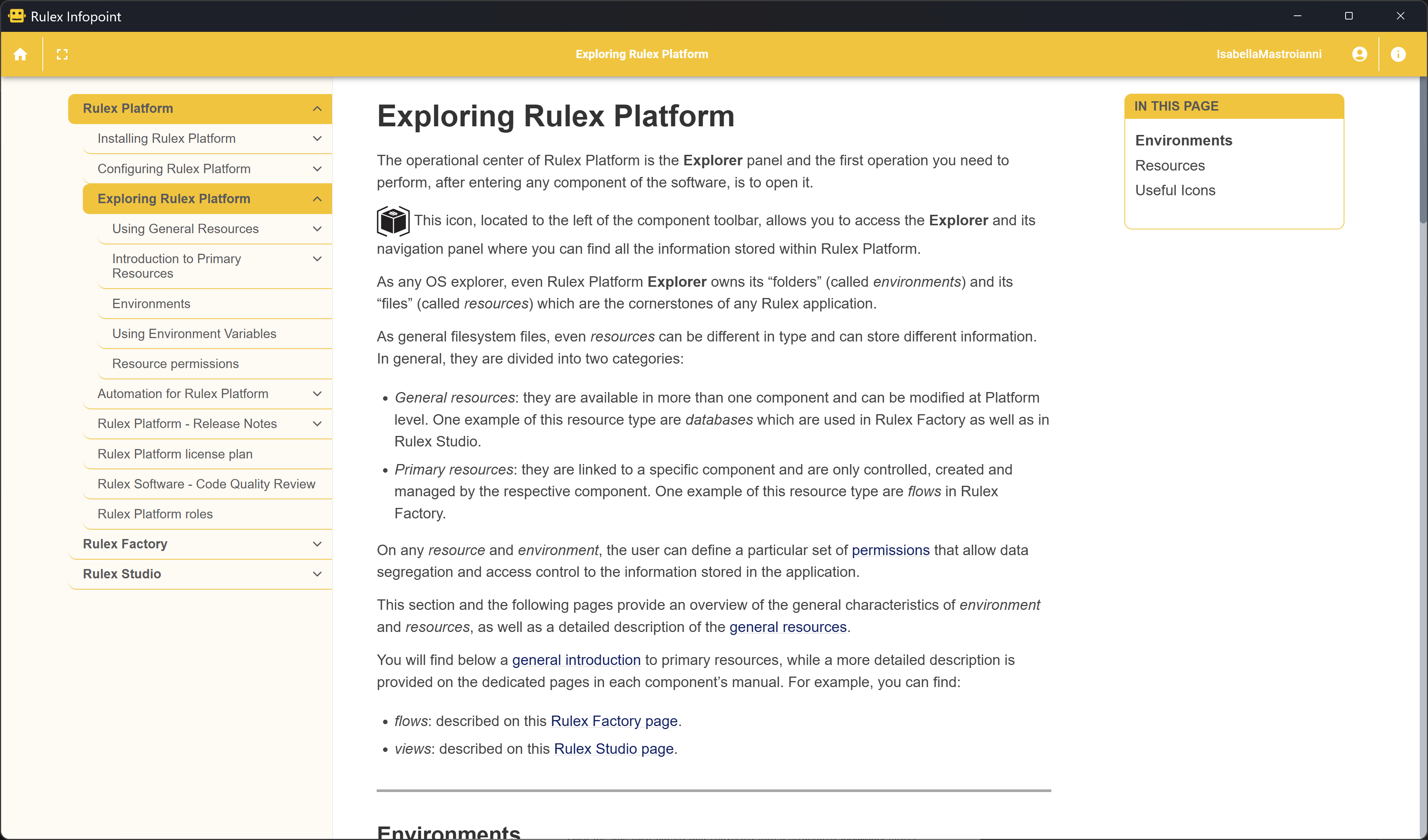}
	\caption{An example of a documentation page of Rulex Platform \cite{rulex_docs_14x}.}
	\label{fig:documentation}
\end{figure}

Each section includes detailed technical explanations and examples of practical procedures, making the content varied in terms of both structure and semantics (Fig. \ref{fig:documentation}). The Rulex Platform Documentation is generated using Sphinx, a tool that converts structured text files into various output formats (e.g., HTML, PDF), while automatically producing cross-references and indexes.

\FloatBarrier
\subsection{Benchmark queries}
\label{appendix:queries}

This appendix reports the full set of twenty benchmark queries used in the comparative evaluation of the baseline and link‑aware RAG systems. The queries were authored by Rulex documentation experts and reflect common information‑seeking patterns observed among Rulex Platform users. They span various areas of the Platform, with a particular focus on the Factory component, addressing topics such as data import and export, installation procedures, versioning operations, Factory configuration, automation, and resource management. As discussed in the Limitations and Future Work section, we are already working on extending the evaluation framework to additional components of the Platform, including Rulex Studio.

\begin{description}[font=\bfseries, leftmargin=1cm, style=nextline,
		itemsep=0.6em,      
		topsep=0.4em,       
		parsep=0pt]         
	
	\item[Impala import]
	Explain how to import data from a table in an Impala Database into a Rulex Factory Flow. Pay attention to explain all the necessary information needed for the connection and the corresponding task option needed for setup.
	
	\item[Excel cleaning]
	Provide a detailed explanation of how to import an Excel file and automatically remove any columns that have more than 50\% missing values.
	
	\item[Rulex install]
	Provide instructions for installing Rulex Platform Desktop and list the required prerequisites.
	
	\item[Data export]
	Explain how to export a data structure to a database.
	
	\item[Reshape merge]
	I need to combine multiple datasets after applying Reshape To Wide. Explain how to ensure that the data are compatible and no duplication occurs.
	
	\item[SharePoint connection]
	How can I set up a custom connection to SharePoint in a Rulex Factory task?
	
	\item[Change repository]
	How can I change the repository connection in a already versioned resource?
	
	\item[Push changes]
	Provide instructions on how to push changes in a versioned resource.
	
	\item[Environment variables]
	I need to create an environment variable to substitute my internal variable with an enterprise value. How can I do it?
	
	\item[Macros]
	Explain how to record an event and how to store it as a macro.
	
	\item[Build/solve configuration]
	How can I set up the configuration file for the Build/Solve task?
	
	\item[Permissions]
	Explain what are permissions and how they affect resources on Rulex Platform.
	
	\item[Dataset for LLM]
	Explain how to prepare a dataset to use the Logic Learning Machine task.
	
	\item[Modify a rule]
	I need to modify a rule generated by an algorithm. How can I do it?
	
	\item[Automation window]
	How can I control the configuration of the Rulex Factory flow computation launch through a dynamic window in Rulex Studio?
	
	\item[Mail alerts]
	How can I setup an alert to send a mail when a task setup is causing an error?
	
	\item[Export to local FS]
	Explain how to export a data structure to a local file system in Rulex Platform Server.
	
	\item[Branch variables]
	What are branch variables? Give me an example of their use.
	
	\item[Outlook integration]
	How can I save a connection to Microsoft Outlook?
	
	\item[Vault resources]
	What are vault resources? How can I use them?
	
\end{description}

\newpage
\section{Detailed results}
\label{appendix:details}

This appendix reports the full set of results that complements the analysis presented in Section~\ref{sec:experimental-results}. In addition, it includes a representative example of the retrieved context to support the qualitative interpretation of the retrieval behaviour. Together, these tables and figures provide a granular view of the process of each RAG configuration across the entire benchmark, allowing us to verify that the trends identified in the main text hold consistently across individual queries and prompts.

\subsection{A representative example of retrieved context}\label{appendix:example}

\begin{figure}[!h]
	\centering
	\includegraphics[width=\textwidth]{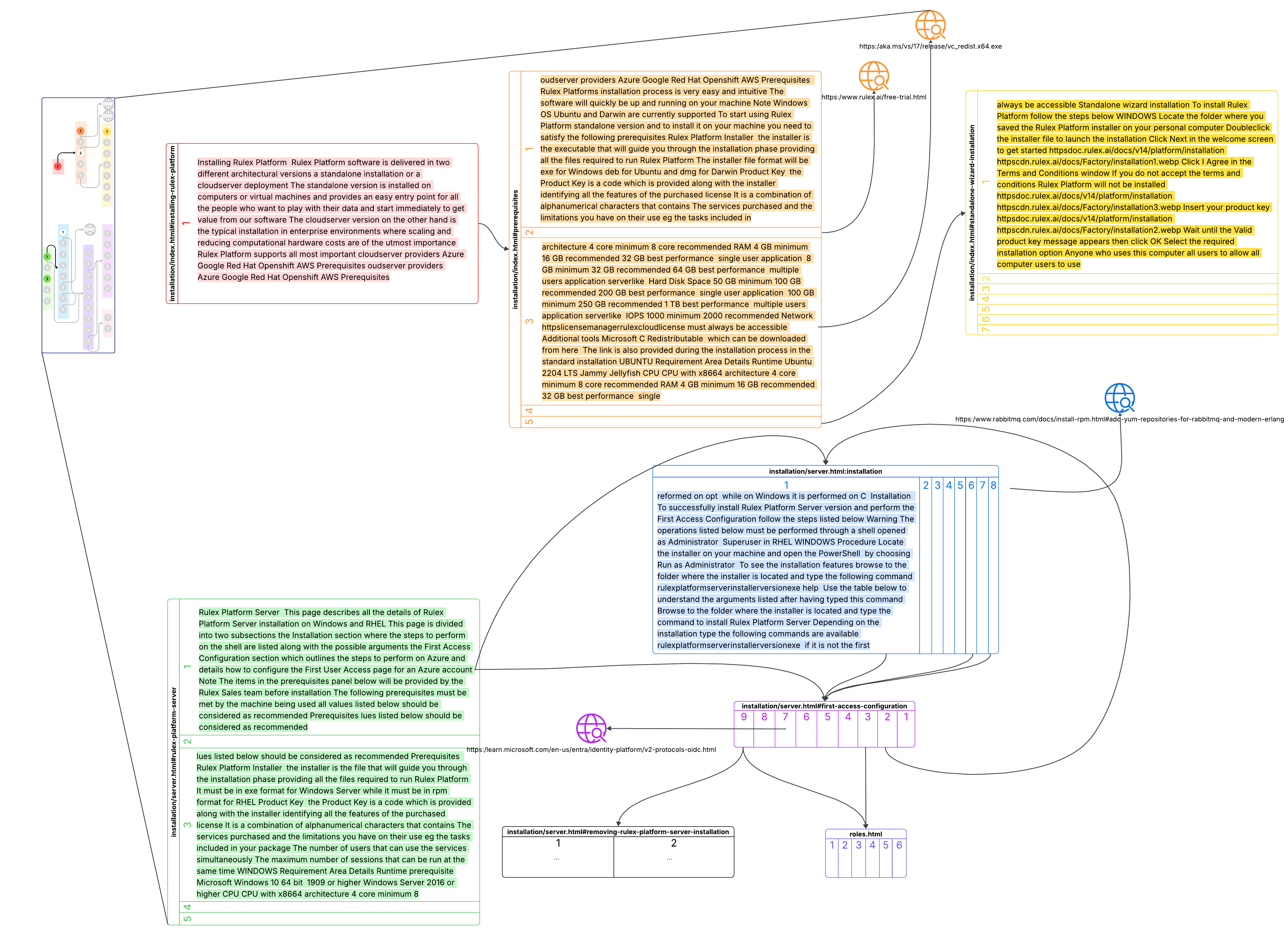}
	\caption{Example of retrieved context for Query~3 under the Basic prompt (see Fig. \ref{fig:k5vsk10vsLA}). The figure shows the full textual content of the chunks retrieved by LARAG, together with the structural relationships induced by hyperlinks.}
	\label{fig:esempiografo}
\end{figure}

Figure \ref{fig:esempiografo} shows in details the context retrieved by LARAG for generating the response reported in Figure \ref{fig:k5vsk10vsLA}.

\FloatBarrier
\subsection{Aggregate results
}\label{appendix:aggr_results}

This section reports aggregate results across all queries, prompts, and models, providing a high-level view of cost-related behaviour and semantic quality before proceeding to more fine-grained per-query and per-prompt analyses.

\FloatBarrier
\subsubsection{Cost metrics}

\begin{figure}[!h]
	\centering
	\includegraphics[width=\textwidth]{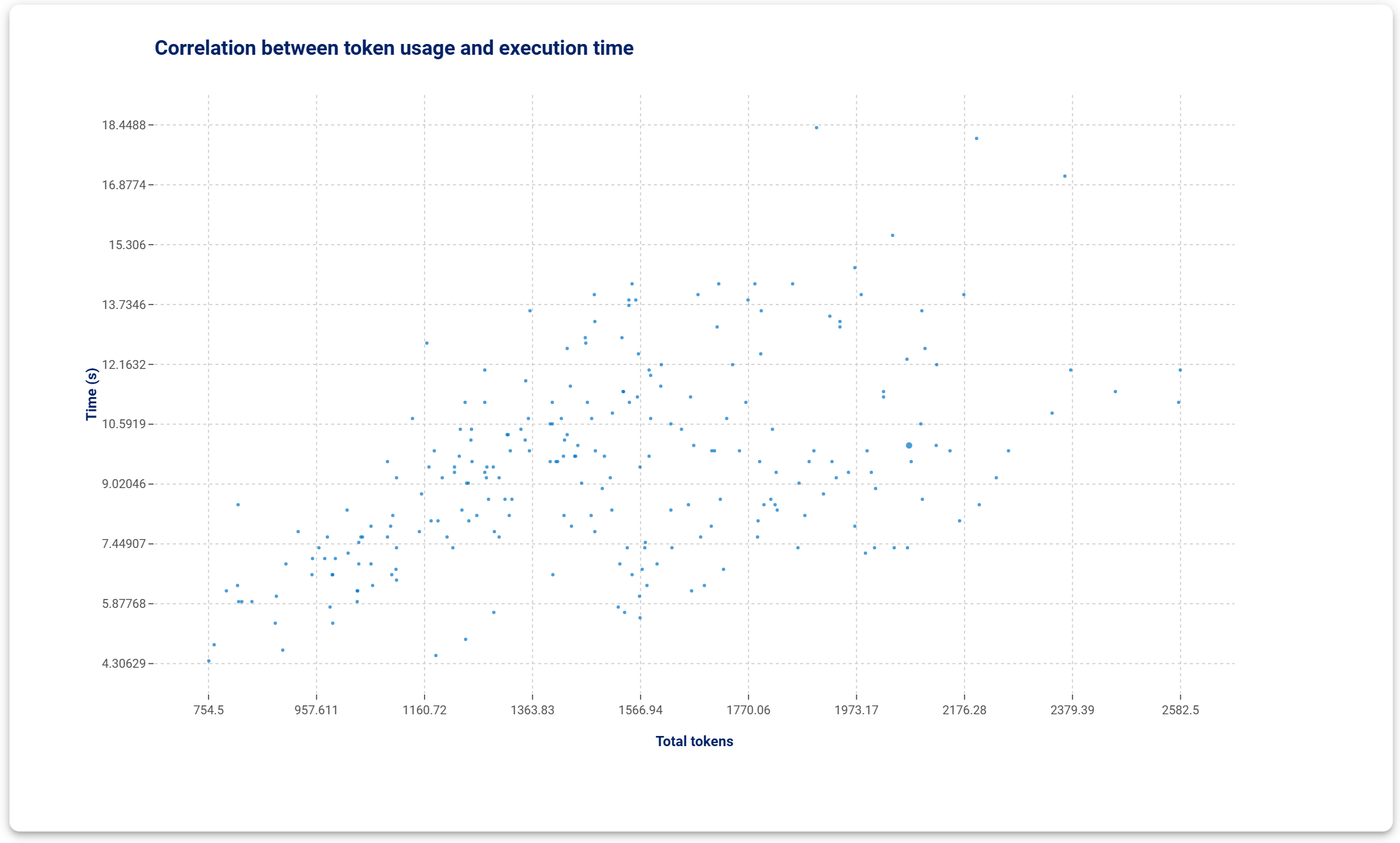}
	\caption{Relationship between total token usage and execution time. Image generated with Rulex Studio.}
	\label{fig:tokens_vs_time}
\end{figure}

Figure~\ref{fig:tokens_vs_time} shows the relationship between total token usage and execution time across all query--prompt pairs. As expected, higher token usage tends to be associated with longer execution times; however, the relationship is far from perfectly linear. Two factors explain why the execution‑time ranking does not align with the token‑usage ranking shown in the figure. First, execution time does not depend solely on the length of the generated answer, but also on the amount and structure of the retrieved context. Queries whose retrieved context is longer or structurally richer require additional attention computation and therefore incur higher latency even when the produced answer is relatively short. This effect is distinct from the gold reference length analysed in Section~\ref{subsec:length_effects}: the latter influences BERTScore, whereas execution time is primarily affected by the size of the retrieved context. Second, some of the residual variability across queries, visible as local inversions between token usage and latency, reflects system-level noise from local execution (e.g. scheduling, temporary load), rather than systematic differences in model behaviour.

%
%
%


\FloatBarrier
\subsubsection{BERTScore}

Table~\ref{tab:bert_global} summarises the global BERTScore means across all models, prompts, and queries. These aggregate values provide a useful reference point for interpreting the more detailed breakdowns discussed later: they capture the overall behaviour of the systems and allow us to verify whether per‑model or per‑query variations deviate from the expected trend. As discussed in the main text, Recall remains consistently higher than Precision, indicating that all configurations tend to prioritise coverage over conciseness. Importantly, this global pattern is preserved when results are disaggregated: none of the models shows systematic deviations from this recall‑driven behaviour in the subsequent analyses.


\begin{table}[h!]
	\centering
	\small
	\begin{tabular}{c c c}
		\toprule
		\textbf{P} & \textbf{R} & \textbf{F1} \\
		\midrule
		0.8296 & 0.8456 & 0.8373 \\
		\bottomrule
	\end{tabular}
	\caption{Global BERTScore means across all queries, prompts, and models.}
	\label{tab:bert_global}
\end{table}

\FloatBarrier
\subsection{Per-query performance}\label{sec:per_query_perf}

This section analyses performance at the level of individual queries, providing a fine-grained view of how different RAG configurations behave across the benchmark. The goal of this analysis is to verify that the aggregate patterns discussed in the main text are consistently observed across the entire benchmark and are not driven by a small subset of queries. A per-query breakdown therefore serves as a robustness check, allowing us to assess whether cost and quality trends remain stable despite variability in query length, complexity, and retrieved context.

\subsubsection*{Cost metrics by query}


\paragraph{Execution time by model.}

\begin{figure}[!h]
	\centering
	\includegraphics[width=0.8\textwidth]{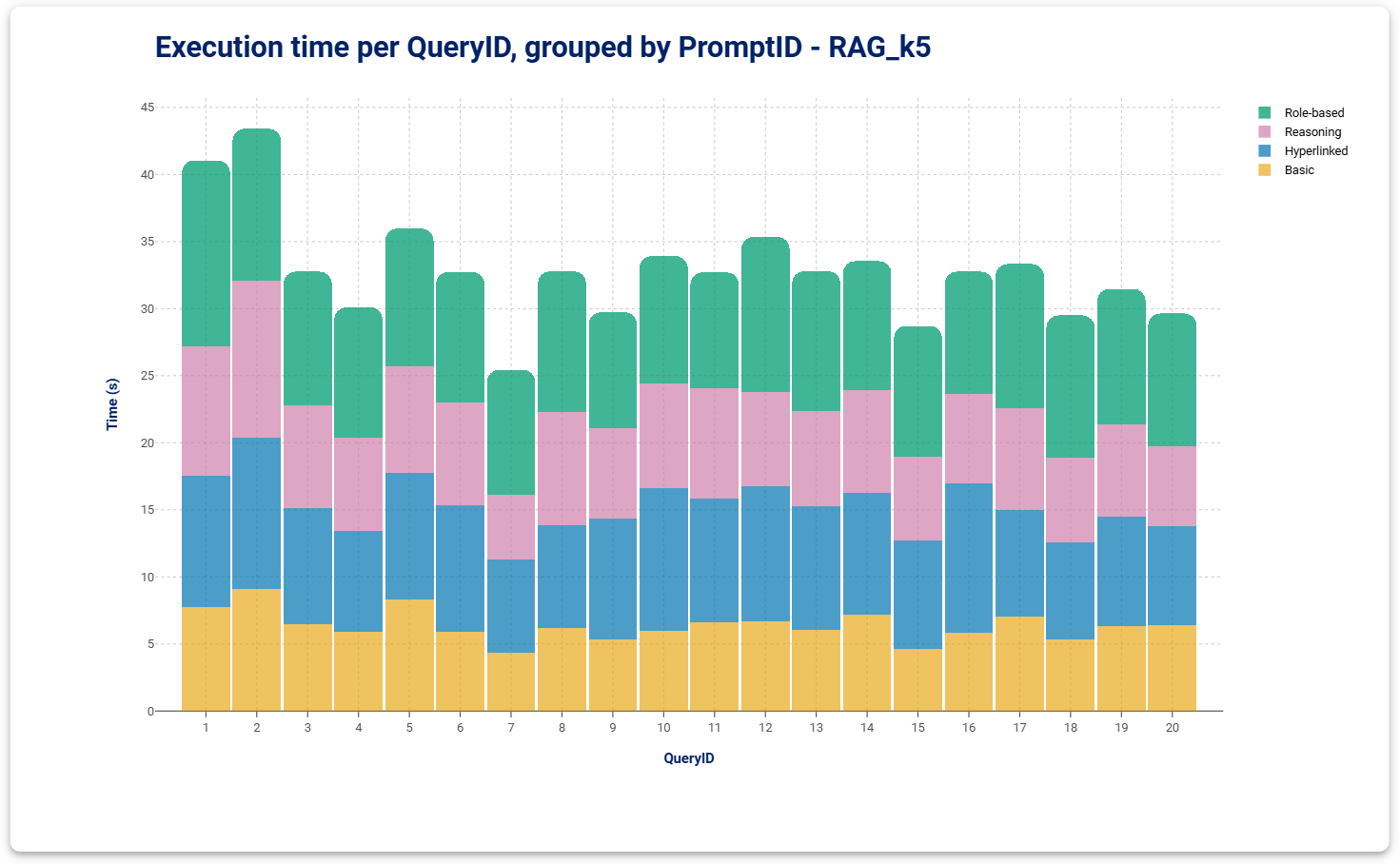}
	\includegraphics[width=0.8\textwidth]{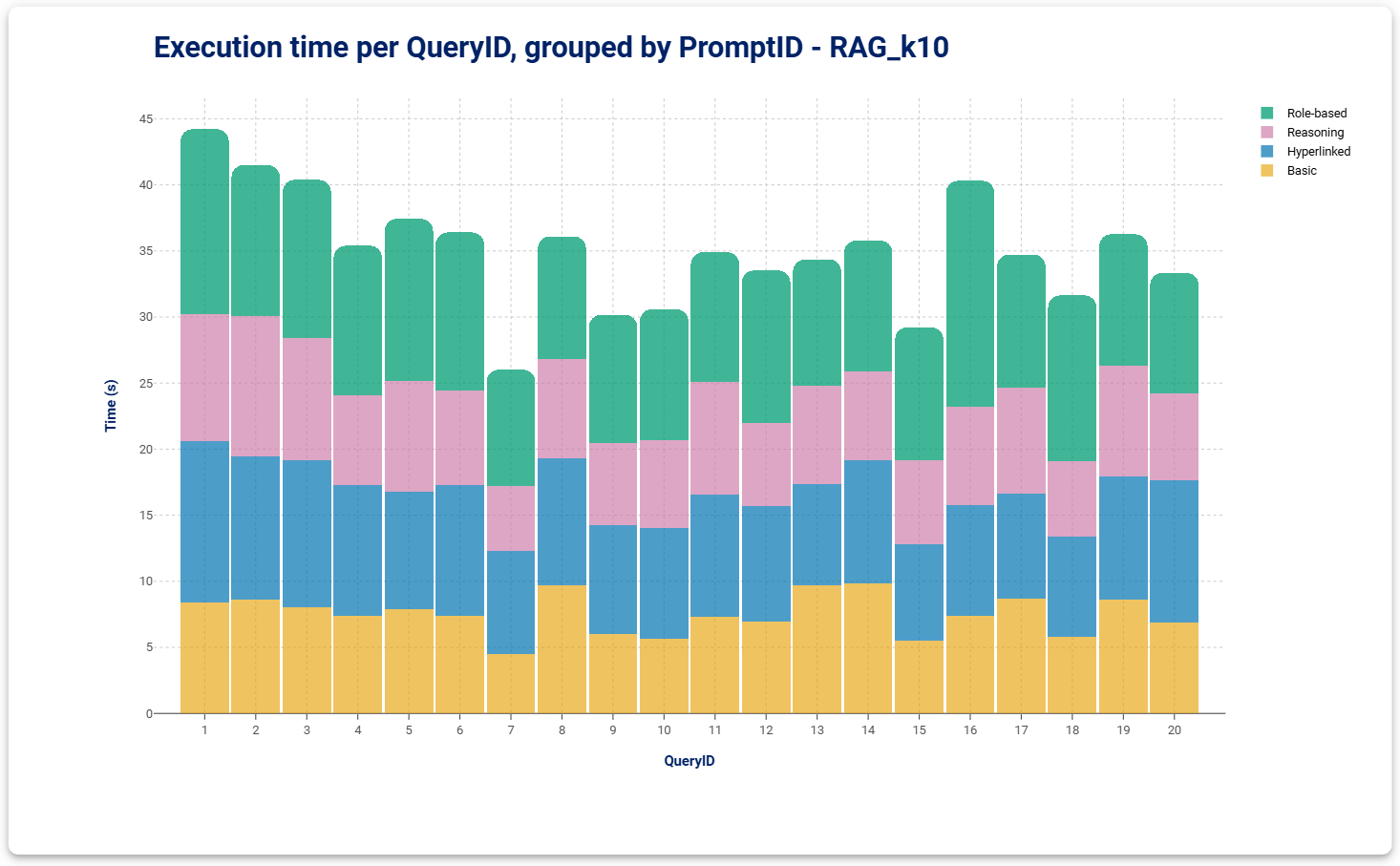}
	\includegraphics[width=0.8\textwidth]{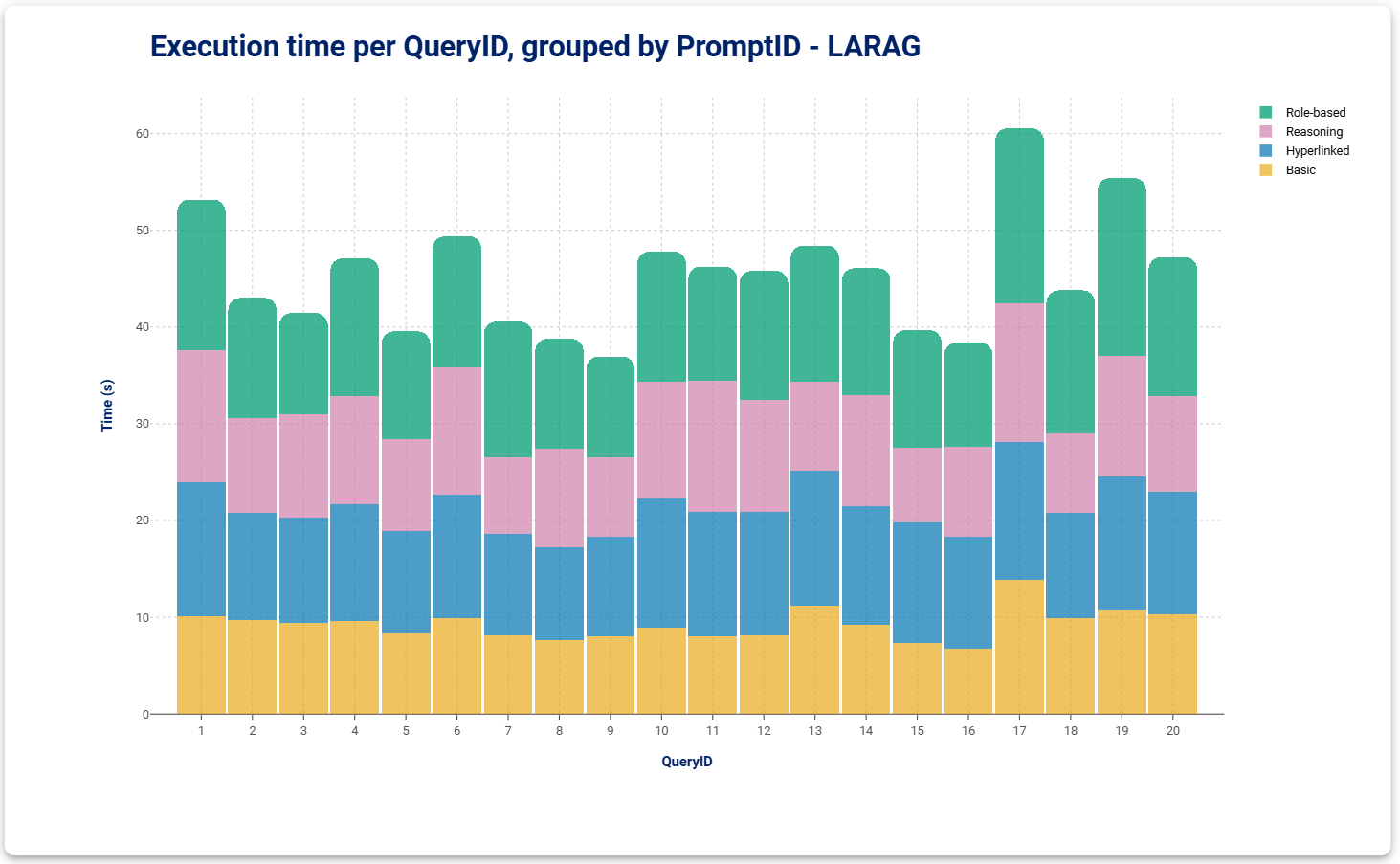}
	\caption{Execution time per each RAG model aggregated by QueryID. Images generated with Rulex Studio.}
	\label{fig:exec_by_model}
\end{figure}

Fig. \ref{fig:exec_by_model} shows execution time aggregated by QueryID separately for RAG\_k5, RAG\_k10, and LARAG. The trends mirror those discussed in the main text: RAG\_k5 is consistently the fastest, followed by RAG\_k10, while LARAG displays query-level variability but remains broadly comparable to RAG\_k10.

%

\paragraph{Token usage by model and prompt.}

\begin{figure}[!h]
	\centering
	\includegraphics[width=0.8\textwidth]{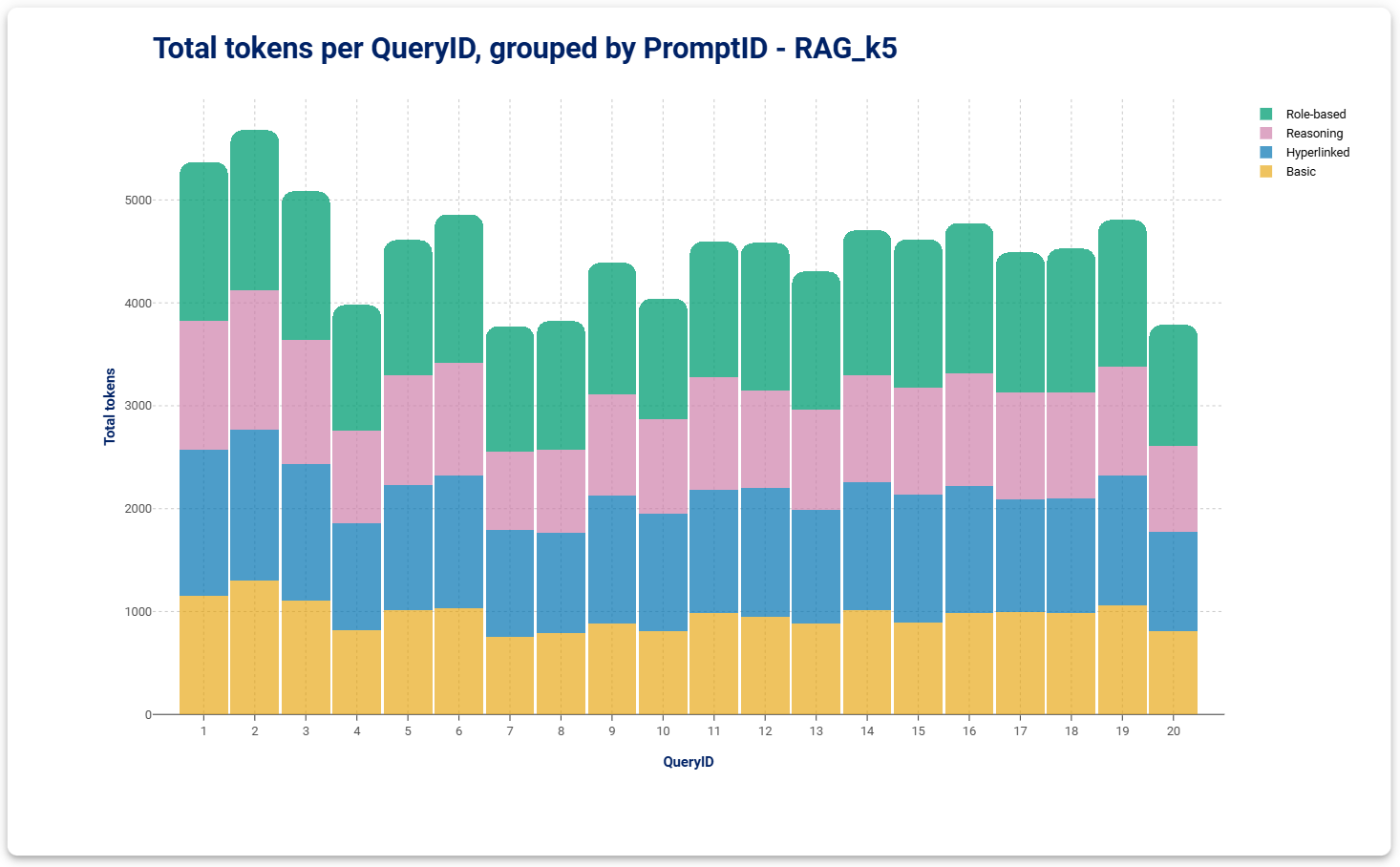}
	\includegraphics[width=0.8\textwidth]{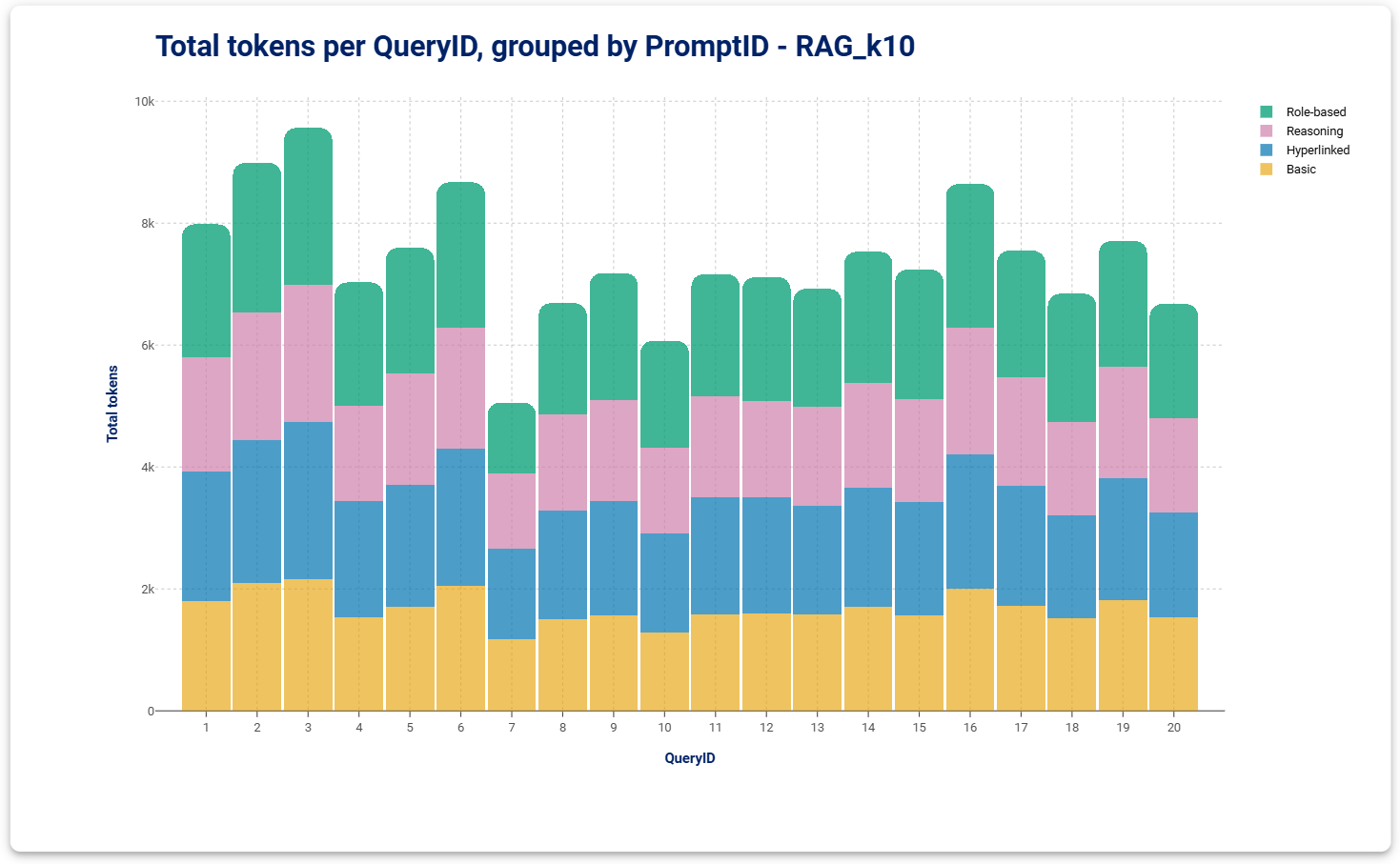}
	\includegraphics[width=0.8\textwidth]{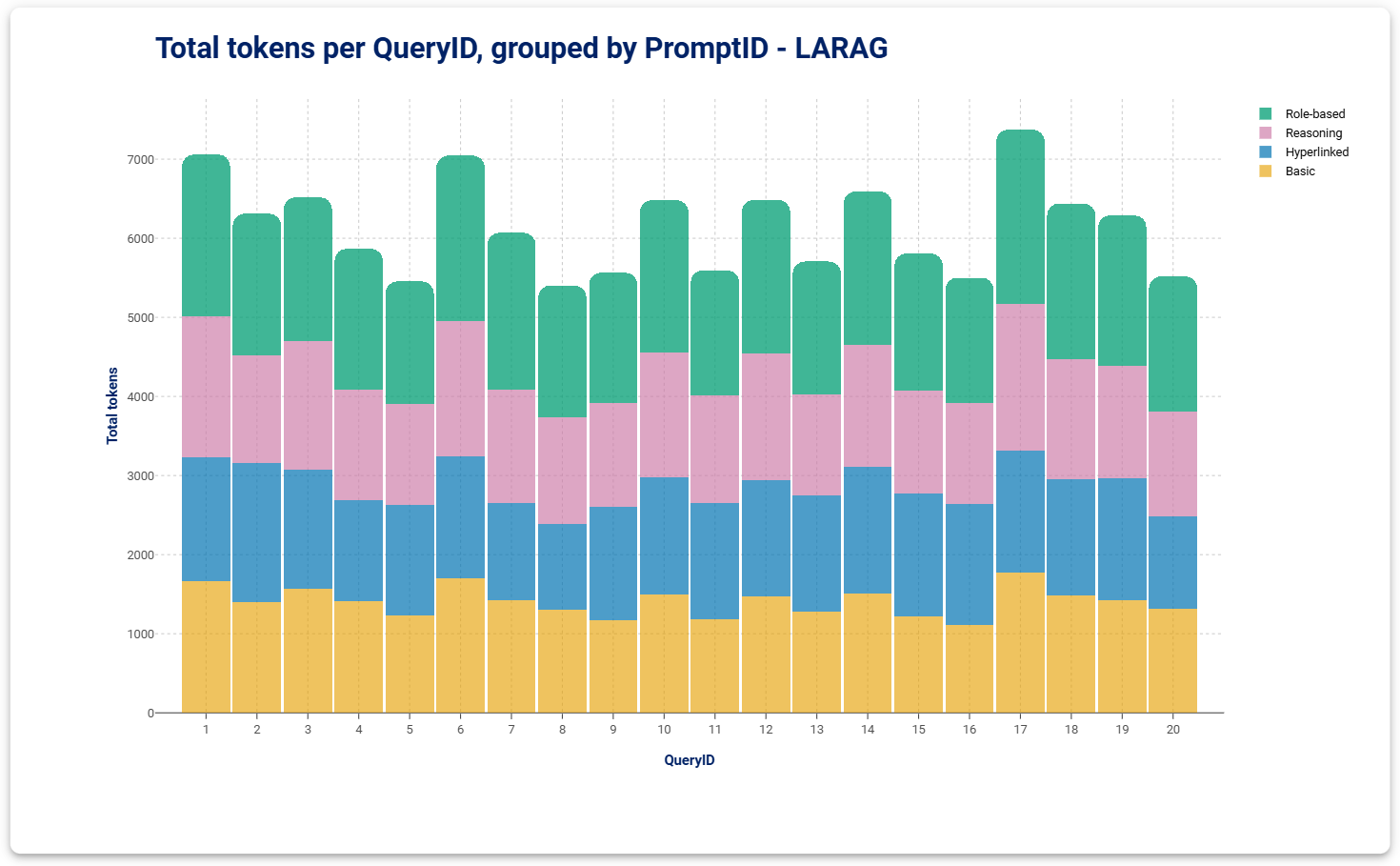}
	\caption{Total tokens per each RAG model aggregated by QueryID. Images generated with Rulex Studio.}
	\label{fig:tokens_by_model}
\end{figure}

Fig.~\ref{fig:tokens_by_model} shows the total number of generated tokens per QueryID, grouped by prompt type. Across all models, Basic prompt produces the fewest tokens, Role-based prompt the most, and Hyperlinked and Reasoning prompt fall in between. This pattern mirrors the execution-time behaviour and confirms that prompting strategies influence cost in a stable manner across queries, as already discussed in Section \ref{sec:experimental-results}.

%

\FloatBarrier
\subsubsection*{BERTScore by query}

Table~\ref{tab:bert_query_all_combined} reports per-query Precision, Recall and F1 for all models. 
A clear pattern emerges: RAG\_k5 frequently attains the highest Precision, a direct consequence of retrieving fewer chunks and therefore introducing less additional content; however, this also limits coverage, resulting in consistently lower Recall and F1. 
RAG\_k10 dominates Recall-oriented queries, while LARAG frequently achieves the best F1 and competitive Precision, especially on longer or multi-step queries. 
These results reinforce the conclusions drawn in Section~\ref{subsec:cost-quality}.


\begin{sidewaystable*}[h!]
	\setlength{\tabcolsep}{2pt} 
	\centering
	\small
	\begin{tabular}{l c c c c c c c c c c c c c c c c c c c c}
		& & & & & & & & & & \textbf{P} & & & & & & & & & & \\
		\toprule
		QueryID: & 1 & 2 & 3 & 4 & 5 & 6 & 7 & 8 & 9 & 10 & 11 & 12 & 13 & 14 & 15 & 16 & 17 & 18 & 19 & 20 \\
		\midrule
		\textbf{RAG\_k5} & 0.8327 & 0.8244 & 0.7968 & 0.8430 & 0.8227 & 0.8260 & \textbf{0.8457} & \textbf{0.8424} & \textbf{0.8467} & 0.8077 & 0.8205 & \textbf{0.8490} & 0.8064 & \textbf{0.8078} & \textbf{0.8288} & 0.8351 & \textbf{0.8344} & 0.8414 & 0.8103 & \textbf{0.8686} \\
		\textbf{RAG\_k10} & \textbf{0.8395} & 0.8253 & 0.7935 & 0.8410 & 0.8227 & \textbf{0.8274} & 0.8378 & 0.8418 & 0.8605 & \textbf{0.8094} & 0.8214 & 0.8485 & \textbf{0.8098} & 0.8075 & 0.8240 & \textbf{0.8357} & 0.8271 & 0.8460 & 0.8126 & 0.8674 \\
		\textbf{LARAG} & 0.8372 & \textbf{0.8330} & \textbf{0.8015} & \textbf{0.8490} &\textbf{ 0.8233} & 0.8205 & 0.8427 & 0.8420 & 0.8431 & 0.8065 & \textbf{0.8259} & 0.8466 & 0.8092 & 0.8055 & 0.8235 & 0.8338 & 0.8251 & \textbf{0.8467} & \textbf{0.8274} & 0.8624 \\
		\bottomrule
	\end{tabular}
	
	\vspace{40pt}
	
	\begin{tabular}{l c c c c c c c c c c c c c c c c c c c c}
		& & & & & & & & & & \textbf{R} & & & & & & & & & & \\
		\toprule
		QueryID: & 1 & 2 & 3 & 4 & 5 & 6 & 7 & 8 & 9 & 10 & 11 & 12 & 13 & 14 & 15 & 16 & 17 & 18 & 19 & 20 \\
		\midrule
		\textbf{RAG\_k5} & 0.8629 & 0.8572 & 0.7997 & 0.8374 & \textbf{0.8330} & 0.8415 & 0.8710 & 0.8650 & 0.8777 & 0.8525 & 0.8262 & 0.8469 & 0.8019 & 0.8333 & 0.8173 & 0.8344 & 0.8557 & 0.8272 & 0.8327 & 0.8495\\
		\textbf{RAG\_k10} & \textbf{0.8718} & 0.8560 & 0.8025 & \textbf{0.8500} & 0.8324 & 0.8472 & \textbf{0.8718} & \textbf{0.8679} & \textbf{0.8803} & 0.8519 & \textbf{0.8304} & 0.8473 & \textbf{0.8048} & 0.8371 & \textbf{0.8182} & \textbf{0.8837} & 0.8586 & 0.8293 & \textbf{0.8465} & \textbf{0.8551} \\
		\textbf{LARAG} & 0.8709 & \textbf{0.8616} & \textbf{0.8270} & 0.8426 & 0.8324 & \textbf{0.8474} & 0.8714 & 0.8664 & 0.8764 & \textbf{0.8548} & 0.8299 & \textbf{0.8526} & 0.8016 & \textbf{0.8374} & 0.8161 & 0.8696 & \textbf{0.8602} & \textbf{0.8321} & 0.8407 & 0.8453 \\
		\bottomrule
	\end{tabular}
	
	\vspace{40pt}
	
	\begin{tabular}{l c c c c c c c c c c c c c c c c c c c c}
		 & & & & & & & & & & \textbf{F1} & & & & & & & & & & \\
		\toprule
		QueryID: & 1 & 2 & 3 & 4 & 5 & 6 & 7 & 8 & 9 & 10 & 11 & 12 & 13 & 14 & 15 & 16 & 17 & 18 & 19 & 20 \\
		\midrule
		\textbf{RAG\_k5} & 0.8475 & 0.8405 & 0.7982 & 0.8402 & 0.8278 & 0.8336 & \textbf{0.8573} & 0.8535 & \textbf{0.8617} & 0.8295 & 0.8233 & 0.8479 & 0.8041 & 0.8203 & \textbf{0.8227} & 0.8525 & \textbf{0.8449} & 0.8341 & 0.8213 & 0.8588 \\
		\textbf{RAG\_k10} & \textbf{0.8553} & 0.8403 & 0.7979 & 0.8453 & 0.8275 & \textbf{0.8372} & 0.8558 & 0.8525 & 0.8605 & \textbf{0.8300} & 0.8258 & 0.8478 & \textbf{0.8073} & \textbf{0.8220} & 0.8210 & \textbf{0.8590} & 0.8425 & 0.8374 & 0.8292 & \textbf{0.8610} \\
		\textbf{LARAG} & 0.8537 & \textbf{0.8471} & \textbf{0.8140} & \textbf{0.8458} &\textbf{ 0.8278} & 0.8337 & 0.8561 & \textbf{0.8540} & 0.8592 & 0.8299 & \textbf{0.8278} & \textbf{0.8495} & 0.8053 & 0.8211 & 0.8195 & 0.8513 & 0.8423 & \textbf{0.8392} & \textbf{0.8337} & 0.8536 \\
		\bottomrule
	\end{tabular}
	
	\caption{Global BERTScore means across all queries, prompts, and models.}
	\label{tab:bert_query_all_combined}
\end{sidewaystable*}

\FloatBarrier
\subsection{Per-prompt performance}\label{sec:per_prompt_perf}

This section focuses on the impact of prompting strategies on both cost and quality metrics. By aggregating results by prompt type, we isolate the contribution of prompt design independently of query content and retrieval configuration. This perspective complements the per-query analysis by highlighting how different prompting choices systematically influence generation length, latency, and semantic similarity with the gold references across the entire benchmark.

\FloatBarrier
\subsubsection*{Cost metrics by prompt}

Table~\ref{tab:cost_prompt} summarises the average cost associated with each prompting strategy. 
The patterns mirror those observed in the main text: Basic prompts are consistently the most efficient; Role-based prompts incur the highest cost due to increased verbosity; Hyperlinked and Reasoning prompts occupy an intermediate position. 
Retrieved chunks remain nearly constant across prompts, indicating that prompting primarily affects the generation phase rather than retrieval depth.

\FloatBarrier
\subsubsection*{BERTScore by prompt}

Table~\ref{tab:cost_prompt} reports BERTScore metrics by prompt type. The results are consistent with the analysis presented in Section~\ref{subsec:prompting}: Basic prompt produces the most efficient outputs, achieving the highest macro F1. Role-based prompt tends to be more verbose and obtain lower BERTScore values due to the inclusion of additional explanatory content, which is not necessarily incorrect and may be only partially captured by similarity-based metrics such as BERTScore, which focus on semantic overlap with a gold reference. Hyperlinked prompt generates more structured responses with stable quality and achieves the highest recall, whereas Reasoning prompt underperforms across all metrics.


\begin{table}[h!]
	\centering
	\small
	\begin{tabular}{c c c c c c}
		\toprule
		\textbf{Prompt} & \textbf{Total tokens} & \textbf{Time (s)} & \textbf{F1} & \textbf{P} & \textbf{R} \\
		\midrule
		Basic & \textbf{1347.38} & \textbf{7.72} & \textbf{0.8444} & \textbf{0.8457} & 0.8434 \\
		Role-based & 1748.83 & 11.53 & 0.8276 & 0.8100 & 0.8462 \\
		Reasoning & 1405.72 &8.53 & 0.8405 & 0.8379 & 0.8435 \\
		Hyperlinked & 1533.08 & 10.05 & 0.8368 & 0.8248 & \textbf{0.8494} \\
		\bottomrule
	\end{tabular}
	\caption{Metrics per prompt.}
	\label{tab:cost_prompt}
\end{table}

It is important to note that this behaviour does not necessarily indicate lower response quality. Similarity-based metrics such as BERTScore focus on semantic overlap with gold references and may therefore penalise answers that introduce useful but non-reference information, such as tips, warnings, or contextual guidance. This effect is also reflected in the higher recall observed for the Hyperlinked prompt, which generates more structured and comprehensive outputs while maintaining stable semantic quality. Reasoning prompt, instead, consistently underperforms across all metrics, confirming its limited effectiveness in this setting.

\FloatBarrier
\subsection{Correlation between length measures and BERTScore}\label{sec:corr_length_BERTScore}

The full correlation matrix between reference length, prediction length, and F1 is reported in Table~\ref{tab:corr_len_f1}. 
As discussed in Section~\ref{subsec:length_effects}, reference length exhibits a much stronger negative correlation with F1 than prediction length, reflecting the greater difficulty of fully covering long references. 
This detailed matrix confirms the weak relationship between \texttt{len\_ref} and \texttt{len\_pred} (0.07), indicating that the two length measures capture distinct aspects of the responses.

\begin{table}[h!]
	\centering
	\small
	\begin{tabular}{lccc}
		\toprule
		& \texttt{len\_ref} & \texttt{len\_pred} & F1 \\
		\midrule
		\texttt{len\_ref}      & 1.000000 & 0.073079 & -0.535856 \\
		\texttt{len\_pred}     & 0.073079 & 1.000000 & -0.334123 \\
		F1  & -0.535856 & -0.334123 & 1.000000 \\
		\bottomrule
	\end{tabular}
	\caption{Correlation matrix between reference length, prediction length, and F1.}
	\label{tab:corr_len_f1}
\end{table}

\medskip
Overall, the detailed results presented in this appendix confirm the robustness of the findings discussed in the main text. Aggregate analyses reveal that cost‑efficiency differences across models and prompts are highly consistent, with RAG\_k5 remaining the fastest, RAG\_k10 trading higher latency for broader coverage, and Link‑A achieving a more favourable efficiency-coverage balance. Per‑query and per‑prompt breakdowns show that these behaviours persist across individual benchmark items, with only minor variability attributable to context size or system‑level noise. The correlation matrix further supports these conclusions by highlighting that reference length, rather than generation length, drives most of the difficulty captured by F1. Taken together, these granular evaluations demonstrate that the trends identified at the aggregate level hold across the full benchmark and reinforce the central claim that hyperlink‑guided retrieval provides quality gains without incurring substantial additional cost.

\end{document}